\documentclass[%
 reprint,
superscriptaddress,
 amsmath,amssymb,
 aps,
prb,
floatfix]{revtex4-2}

\usepackage{chemformula}
\usepackage{graphicx}%
\usepackage{subcaption} %
\usepackage{dcolumn}%
\usepackage{placeins} %
\usepackage{bm}%
\usepackage[colorlinks, allcolors=blue]{hyperref}
\usepackage{multirow}
\usepackage{subcaption} 
\usepackage[export]{adjustbox} 
\usepackage{cleveref} 
\usepackage{physics}
\usepackage{amsmath}
\usepackage{color}

\usepackage{rotating}
\usepackage{booktabs}

\begin{document}

\title{%
Evaluating covalency using 
RIXS spectral weights:
Silver fluorides vs. cuprates
}%

\author{Ilya Degtev}
\affiliation{ISC-CNR, Istituto dei Sistemi Complessi, via dei Taurini 19, 00185 Rome, Italy}%
\affiliation{Dipartimento di Fisica, Sapienza Università di Roma, 00185 Rome, Italy}

\author{Daniel Jezierski}
 \affiliation{University of Warsaw, Center of New Technologies, Żwirki i Wigury 93, 02-089 Warsaw, Poland}

\author{Adri\'{a}n G\'{o}mez Pueyo}
\affiliation{ISC-CNR, Istituto dei Sistemi Complessi, via dei Taurini 19, 00185 Rome, Italy}%
\affiliation{Dipartimento di Fisica, Sapienza Università di Roma, 00185 Rome, Italy}

\author{Luciana Di Gaspare}
\affiliation{Dipartimento di Scienze, Università Roma Tre, 
Viale Guglielmo Marconi 446, 00146 Rome, Italy}

\author{Monica De Seta}
\affiliation{Dipartimento di Scienze, Università Roma Tre, 
Viale Guglielmo Marconi 446, 00146 Rome, Italy}

\author{Paolo Barone}
\affiliation{SPIN-CNR, Istituto Superconduttori, Materiali Innovativi e Dispositivi, Area della Ricerca di Tor Vergata, via del Fosso del Cavaliere 100, 00133 Rome, Italy}	

\author{Giacomo Ghiringhelli}
\email{giacomo.ghiringhelli@polimi.it}
\affiliation{Dipartimento di Fisica, Politecnico di Milano, piazza Leonardo da Vinci 32, 20133 Milano, Italy} \affiliation{CNR-SPIN, Dipartimento di Fisica, Politecnico di Milano, piazza Leonardo da Vinci 32, 20133 Milano, Italy} 

\author{ Pieter Glatzel }
\affiliation{ESRF – The European Synchrotron, 71, avenue des Martyrs, 38000 Grenoble, France}

\author{Zoran Mazej}
 \affiliation{%
Jožef Stefan Institute, Department of Inorganic Chemistry and Technology, Jamova cesta 39, 1000 Ljubljana, Slovenia
}%

\author{Wojciech Grochala}%
\email{w.grochala@cent.uw.edu.pl}
\affiliation{University of Warsaw, Center of New Technologies, Żwirki i Wigury 93, 02-089 Warsaw, Poland}

\author{Marco Moretti Sala}
\email{marco.moretti@polimi.it}
\affiliation{Dipartimento di Fisica, Politecnico di Milano, piazza Leonardo da Vinci 32,
20133 Milano, Italy}

\author{Jos\'{e} Lorenzana}
\email{jose.lorenzana@cnr.it}
\affiliation{ISC-CNR, Istituto dei Sistemi Complessi, via dei Taurini 19, 00185 Rome, Italy}%
\affiliation{Dipartimento di Fisica, Sapienza Università di Roma, 00185 Rome, Italy}

\date{\today}%

\begin{abstract}
We investigate the electronic structure of \ch{AgF2},  \ch{AgFBF4}, \ch{AgF} and \ch{Ag2O}  using X-ray absorption spectroscopy (XAS) and resonant inelastic X-ray scattering (RIXS) at the Ag $L_3$ edge. XAS results were compared with density functional theory computations of the spectra, allowing an identification of main features and an assessment of the theoretical approximations. Our RIXS measurements reveal that \ch{AgF2} exhibits charge transfer excitations and $dd$ excitations, analogous to those observed in \ch{La2CuO4}.  We propose to use the ratio of $dd$ to CT spectral weight as a measure of the covalence of the compounds and provide explicit equations for the weights as a function of the scattering geometry for crystals and powders. The measurements at the metal site $L_3$ edge and previous measurement in the ligand $K$ edge reveal a striking similarity between the fluorides and cuprates materials, with fluorides somewhat more covalent than cuprates.  These findings support the hypothesis that silver fluorides are an excellent platform to mimic the physics of cuprates, providing a promising avenue for exploring high-$T_c$ superconductivity and exotic magnetism in quasi-two-dimensional (\ch{AgF2}) and quasi-one-dimensional (\ch{AgFBF4}) materials. 
\end{abstract}

\maketitle

\section{Introduction}

In the search for superconducting materials similar to cuprates, a natural route is the replacement of Cu by Ag, as they both belong to the coinage metals group. Despite this chemical similarity, copper and silver have important differences. Because silver is more electronegative than copper, \ch{AgO} has a negative charge transfer energy that ends up in a non-magnetic charge disproportionated state (1+/3+), instead of the desired $4d^9$ \ch{Ag^{2+}}. This is in contrast to \ch{CuO}, which is an antiferromagnetic charge transfer insulator with the $3d^9$ ground state of \ch{Cu^{2+}}.  The lack of magnetism in \ch{AgO} can be addressed by replacing each oxygen with two fluorine atoms, resulting in \ch{AgF2}. 
The stronger electronegativity of fluorine compensates for the electronegativity of Ag and results in the wished spin-1/2 antiferromagnetic charge-transfer insulator that has been proposed \cite{Gawraczynski2019,Miller2020} as an alternative route to high-temperature superconductivity.

\begin{figure}[t]
\centering
\includegraphics[width=0.5\textwidth]{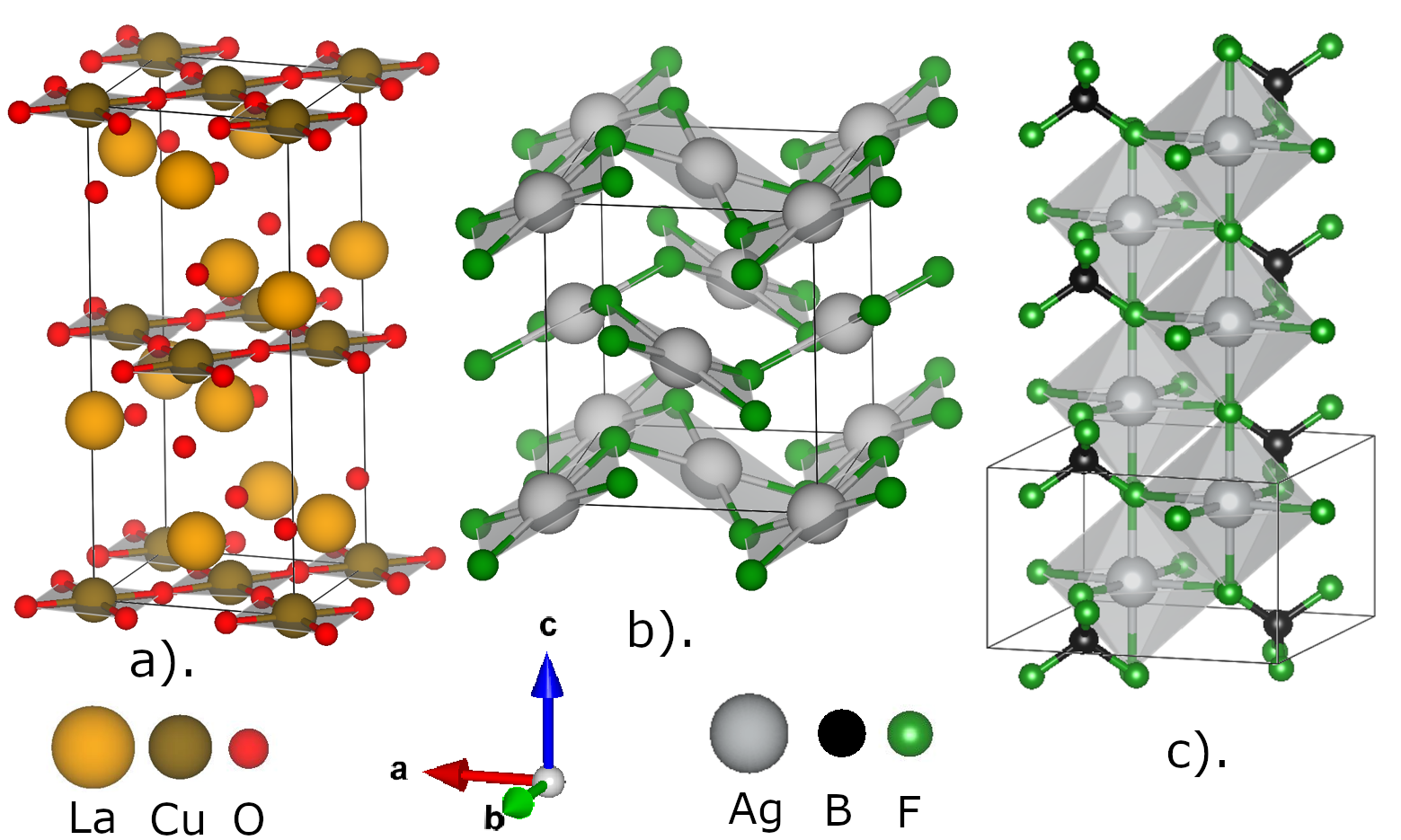}
\caption{\label{fig:AgF2_structure} Crystallographic structures of \ch{La2CuO4}(a), \ch{AgF2}(b) and \ch{AgFBF4}(c).  
}
\end{figure} %

Unlike the charged \ch{CuO2} plane of cuprates [Fig.~\ref{fig:AgF2_structure}(a)], \ch{AgF2} is inherently neutral due to the presence of \ch{F-} ions instead of \ch{O^{2-}}. Consequently, \ch{AgF2} does not need a charge reservoir [Fig.~\ref{fig:AgF2_structure}(b)] and can be regarded as the ``012" equivalent of the 214 stoichiometry found in cuprates like $\ch{La2CuO4}$. 
Indeed, \ch{AgF2} consists of neutral planes stacked one on top of each other, which have the same topology as the \ch{CuO2} planes of cuprates but with much larger buckling [c.f. Fig.~\ref{fig:AgF2_structure} (a) and (b)].

The magnetic ground state of \ch{AgF2} has been determined long ago \cite{Fischer1971I}. It is a canted antiferromagnet with a weak ferromagnetic component larger than the analogous phenomena in parent cuprates. This can be understood as due to the larger buckling of planes 
of \ch{AgF2} and larger spin-orbit coupling leading to a stronger Dzyaloshinskii–Moriya interaction. 

The Néel temperature \cite{Fischer1971I} of \ch{AgF2} (163 K) is half that of \ch{La2CuO4} (325 K) \cite{Kastner1998}, which hints at a moderately large antiferromagnetic interaction.
Indeed, the antiferromagnetic interaction has been measured with two-magnon Raman scattering and inelastic neutron scattering\cite{Gawraczynski2019}, yielding a value $J=70$~meV compared to $J=100 \sim 130$~meV in the Cu-based siblings. 

In cuprates, theoretical \cite{Keimer2015} as well as
empirical arguments suggest that spin fluctuations are important for Cooper pairing with a reported positive correlation between $T_c$ and doping for monolayer systems \cite{Moreira2001,Ofer2006, Grzelak2020}. 

It has been proposed \cite{Grzelak2020} that the antiferromagnetic interaction of \ch{AgF2} 
could be enhanced by growing a flat layer on appropriate substrates. This, in addition, has the advantage of reducing polaronic tendencies, which could hamper metallicity \cite{Bandaru2021}.  This strain engineering approach predicts a maximum superconducting $T_c$ of nearly 200~K. However, it relies on a close analogy between \ch{AgF2} and parent cuprates, which calls for an in-depth comparison.

Although optical measurements are challenging due to the lack of single crystals, strongly compressed powder samples have been produced using explosives, which allowed the recording of optical spectra \cite{Bachar2022}. The optical absorption shows an onset at 1.75~eV for charge-transfer excitations and a broad peak at 3.4 eV. These energies are somewhat higher than, for example, \ch{La2CuO4} (LCO), which shows an onset at about 1.6~eV and a peak at 2.2~eV.  Hybrid density functional theory (DFT) calculations predicted a charge transfer parameter $\Delta=2.7$~eV for \ch{AgF2} \cite{Gawraczynski2019}.
High-energy spectroscopies combined with small cluster exact computations \cite{Piombo2022} yield a somewhat higher value $\Delta=3.5\sim 4.1$~eV.
 
In addition to optical spectroscopy, 
Ref.~\cite{Bachar2022} presented the resonant inelastic X-ray scattering (RIXS) spectrum of \ch{AgF2} at the fluorine $K$ edge  (682 eV, soft X-rays). The spectrum is dominated by charge transfer excitations above 3 eV, but also $dd$ excitations around 2 eV are clearly visible due to mixing between the transition metal and the ligand. 
Moreover, a tail of the elastic peak around 200 meV has been assigned to bimagnon excitations. Interestingly, the spectrum of \ch{AgF2} holds several analogies to that of LCO measured at the O $K$ edge. This is a very encouraging result for attempting to dope and search for superconductivity in \ch{AgF2}. Because the  RIXS was performed on the F $K$ edge and on a powder sample, polarization information was not available. Therefore, the assignment of the symmetry of the different $dd$ excitations could be done only indirectly by comparing the energy with DFT computations.

Besides the potential interest as cuprate analogues\cite{Yang2014,Yang2015,Gawraczynski2019,Miller2020,Grzelak2020,Rybin2021}, silver fluorides are interesting because of the possibility to host exotic magnetic phases \cite{McLain2006,Zhang2011,Kurzydowski2013,Kurzydowski2017Prediction,Kurzydowski2017Large,Kurzydowski2018,Sanchez-Movellan2021,Tokar2021,Koteras2022,Prosnikov2022,Wilkinson2023}. In this regard, a particularly intriguing system is \ch{AgFBF4} [Fig.~\ref{fig:AgF2_structure}(c)]. Initially thought to be a metal due to its temperature-independent susceptibility \cite{Casteel1992}, it is now believed to be an excellent realization of the one-dimensional Heisenberg model \cite{Kurzydowski2017Prediction,Kurzydowski2017Large} with a superexchange interaction of $J\approx 300$~meV. Ironically, this model can be mapped to a one-dimensional ``metallic"  system, but in which the fermionic excitations are not electrons but neutral spinon excitations \cite{Giamarchi2004}. 

According to DFT computations \cite{Kurzydowski2017Prediction,Kurzydowski2017Large},  the strong superexchange in \ch{AgFBF4} is due to partially filled $d_{3z^2-r^2}$ orbitals oriented along the $c$-axis hybridizing with ligand $p_z$ orbitals. 
Up to now, one of the best realizations of the one-dimensional Heisenberg model is \ch{Sr2CuO3}, whose spinon spectrum observable in optics, can be perfectly fitted with this model \cite{Lorenzana1997} with $J=246$~meV.  The spectrum has a logarithmic singularity at $\omega_s=J\pi/2=386$~meV, which could be observed in RIXS \cite{Schlappa2012}, and the extracted value of $J$ coincides with the theoretical fit of the optical experiments. In the case of \ch{AgFBF4}, which we study here, the value of $J$ predicted by DFT implies a Van Hove singularity in the RIXS spectrum (due to powder average) at $\omega_s$=470~meV.

\begin{table*}[t]
\caption{\label{tab:conv_param_xas}
Parameters used in the DFT computations. The standard VASP parameters are: Energy cutoff for the plane-wave basis (ENCUT), parameters for the Monkhorst-Pack (MP) $k$-point grid, and total number of states (NBAND).
}
   \begin{ruledtabular}
    \centering
    \begin{tabular}{c|cccccccc}
        \multirow{2}{*}{} & \multicolumn{3}{c}{Lattice Constants} &ENCUT   & \multicolumn{2}{c}{XAS} &\multicolumn{2}{c}{DOS} \\
\cmidrule(lr){2-4} 
\cmidrule(lr){6-7} \cmidrule(lr){8-9}
      Compound   &  $a$ (\AA)& $b$ (\AA)& $c$ (\AA) &(eV) & MP $k$-point grid & NBAND &  MP $k$-point grid & NBAND \\
        \midrule
    Ag       & 4.052 & 4.052 & 4.052 & 700 & $15\times 15\times 15$ & 1504  & $48\times 48\times 48$ & 150  \\
  \ch{Ag2O}  & 4.716 & 4.716 & 4.716 & 600 & $9\times 9\times 9$    & 300   & $15\times 15\times 15$ & 150  \\
  AgF        & 4.900 & 4.900 & 4.900 & 700 & $13\times 13\times 13$ & 360   & $19\times 19\times 19$ & 150  \\
  \ch{AgF2}  & 5.073 & 5.529 & 5.813 & 600 & $10\times 10\times 10$ & 500   & $16\times 16\times 16$ & 200  \\
 \ch{AgFBF4} & 6.670 & 6.670 & 3.995 & 600 & $9\times 9\times 15$   & 504   & $15\times 15\times 25$ & 200  \\
    \end{tabular}
 \end{ruledtabular}   
\end{table*}

For insulating cuprates, the determination of $dd$ excitations\cite{Ghiringhelli2004,MorettiSala2011}, and the full characterization of the single and multimagnon  spectra\cite{Braicovich2009,Dean2012,Peng2017,Betto2021} were made by Cu $L_3$ RIXS. This is because at the $L_3$ edge (930 eV), those excitations are accessed directly and not indirectly as at the ligand $K$ edge. Indeed, in $3d$ transition metals, the $2p\rightarrow 3d$ resonance is particularly intense, providing ideal conditions for RIXS spectroscopy. 
 
Even though \ch{AgF2} is not a new material, advanced spectroscopy studies are scarce, due to its instability and the lack of single crystals. To have a complete comparison of the electronic structure of the \ch{AgF2} and cuprates, it is highly desirable to perform RIXS at the absorption edge of the metal.  Unfortunately, the Ag $L_3$ edge is at 3350~eV (tender X-rays), and the soft and hard X-ray beamlines and high-resolution spectrometers developed for 3$d$ and $5d$ transition metal $L$ edges are unsuitable. 

In order to perform a RIXS measurement at the Ag $L_3$ edge, we used the ID26 Tender X-ray Emission Spectrometer (TEXS) beamline (ESRF, Grenoble) \cite{Rovezzi2020}. Although this beamline is dedicated mainly to high-energy-resolution fluorescence detection and does not have the resolution of present-day experiments at dedicated RIXS beamlines, we show here that RIXS measurements at the Ag $L_3$ edge are feasible, and clear conclusions about the electronic similarity between silver fluorides and cuprates can be drawn. 

The paper is organized as follows. Section~\ref{sec:methods} presents the experimental and DFT methods. In Sec.~\ref{sec:xray} we present silver $L_3$ ($2p\rightarrow 4d$)  X-ray absorption spectroscopy (XAS) measurements of powder samples of \ch{AgF_2} and \ch{AgFBF4} ($d^9$), and, as a reference,  AgF and \ch{Ag2O} ($d^{10}$). 
We compare the spectra with previous literature results, including pure silver, and with DFT computations.
The near-edge spectra of the $d^9$ materials exhibit pre-edge features, indicating the feasibility of RIXS processes. Indeed, a rich RIXS spectrum was obtained at the Ag $L_3$ edge for \ch{AgF2} and \ch{AgFBF4}, while \ch{AgF} and \ch{Ag2O} show essentially only fluorescence signals (Sec.~\ref{sec:rixs}). In Sec.~\ref{sec:Comparison} we compare cuprates and silver fluorides in ligand and metal edges. In Sec.~\ref{sec:covalency}, we present a method to determine covalency from spectral weight ratios and apply it to cuprates and silver fluorides. To this aim, Appendix~\ref{app:weights}
presents general formulae to compute the powder average of RIXS scattering intensities. Finally, we conclude in Sec.~\ref{sec:conc}.

\section{\label{sec:methods}  Methods
}
\subsection{Samples}

We measured polycrystalline samples of \ch{AgF2} and \ch{AgFBF4}. 
Because the samples are very sensitive to moisture, it was necessary to mount them in a specially designed cell with an inert atmosphere and a Be window. This avoided degradation in the low vacuum ($10^{-5}$ mbar) of the TEXS instrument.  \ch{Ag2O} and \ch{AgF} samples were also powder samples.

\subsection{Experimental} 
\label{sec:method_exp}

The measurements were done at beamline ID26 of ESRF -- The European Synchrotron (Grenoble, France) using the Tender X-ray Emission Spectrometer (TEXS) \cite{Rovezzi2020}. The spectrometer can host up to 11 cylindrically bent Johansson crystal analyzers arranged in a non-dispersive Rowland circle geometry and a sixteen-wire detector. For the present experiment, we used one Si(220) Johansson crystal with bending radius of the crystal planes of 1004 mm at a horizontal scattering angle around 96$^\circ$. All the measurements were done with the incoming polarization in the scattering plane ($\pi$ polarization). The incident beam energy was selected by the Si(111) reflection of a cryogenically cooled double crystal monochromator. The combined energy bandwidth as measured in the elastic scattering peak from a Cu holder was 0.5 eV. All measurements were performed at $T = 15$~K. 

\begin{figure}[tb]
\centering
\includegraphics[width=\columnwidth]{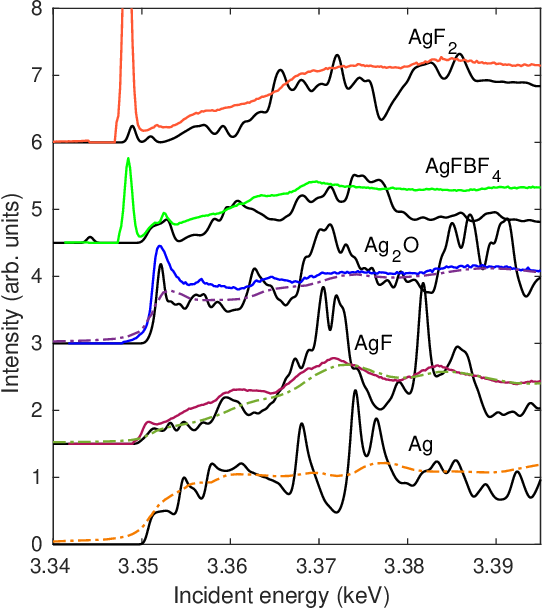}
\caption{\label{fig:XANES} 
Absorption spectra of the investigated compounds and silver in the region of the Ag $L_3$ absorption edge.
We show data from the present work (full colored lines) and Ref.~\cite{Miyamoto2010} (dot-dashed lines). The curves were normalized to 1 at the highest energy measured and plotted with a vertical shift of 1.5 (arb. un.).
We also show the SCH results (black).  The intensity of the DFT data was adjusted to ease the comparison with the experiment. 
The theoretical spectra were shifted by 2 eV to higher energy,  except for \ch{AgF2} (+4~eV shift) and Ag (-2~eV shift).
}
\end{figure}

We used high-energy-resolution fluorescence-detected (HERFD) XAS. This technique\cite{Jaklevic1977,Carra1995,Hamalainen1991,Orduz2024} consists of measuring the 
X-ray absorption near-edge structure (XANES)
of the incident photons by monitoring the intensity of a fluorescence line, in our case L$\beta_{2,15}$, using a narrow energy resolution. 
The L$\beta_{2,15}$ line corresponds to the decay $4d\rightarrow 2p_{3/2}$. It should be noted that the fluorescence line used for detection is in the same energy region as the outgoing RIXS photons. The L$\beta_{2,15}$ line was chosen for HERFD-XANES instead of L$\alpha_{1}$ in order to avoid moving the XES instrument over large angular ranges between XANES and RIXS measurements. As a consequence, the elastic peak appears in the HERFD-XANES scans at the energy that was chosen for the emission spectrometer (3346.6 eV). In the XAS spectra shown in Fig.~\ref{fig:XANES}, we removed the elastic line from the data for the sake of clarity. 

Regarding RIXS, a two-dimensional map is constructed by scanning the energies of both incoming and outgoing photons. Due to the limitation in resolution, the quasielastic line likely has significant inelastic contributions (for example, from magnetic excitations and phonons). Thus, the energy of our ``elastic" feature has a systematic positive error, which implies an underestimation of the reported energy of excitations (of the order of the energy resolution) as discussed below.  The count rates in the $dd$-excitations were tens of kHz.

Several scans were performed for each sample to check for possible degradation during measurement. In the absence of significant changes, the reported spectra are the average over such scans. Despite the low energy resolution, as we shall see, the main structures could be identified and compared to typical spectra in cuprates. 

\subsection{DFT computations}

We used the Vienna Ab initio Simulation Package~\cite{VASP1,VASP2,VASP3,Kresse1999} (VASP) to compute the XAS functions of bulk \ch{AgF2} and related compounds using the supercell core-hole (SCH) method\cite{Karsai2018} with the PBEsol~\cite{Perdew2008} exchange and correlation functional. In the case of $d^9$ compounds, we repeated the computation in the antiferromagnetic state using the DFT+$U$ corrections with $U=5$ eV. For the crystal structures, we used the experimental data.

In the following, VASP parameters are indicated by their standard name in capital letters.  We used Gaussian smearing with a width given by SIGMA=0.02 eV, while for the XAS spectrum, the corresponding smearing parameter was set at CH\_SIGMA=0.5~eV. The self-consistent loop was interrupted when the error in the energy was less than EDIFF=$10^{-6}$ eV. VASP parameters used for the energy cutoff of the plane-wave basis and the Monkhorst-Pack grid are detailed in Table~\ref{tab:conv_param_xas} for each compound, together with the cell dimensions.

In the SCH method, the absorption is computed in the presence of a real-space core-hole. To minimize the effect of the periodic images of the core hole, one can perform the computation in a supercell. 
We checked that for \ch{AgF2} the results were almost identical in the unit cell (4 Ag atoms) or in a  $2\times 2\times 2$ supercell. Therefore, for the fluorides and the oxide, we performed the computation in the unit cell, as we do not expect they would behave differently. 
For Ag we found that the elementary cell was not enough, and we used a $2\times 2\times 2$ supercell.  %

A window of about 50~eV above the Fermi level is necessary to compare with the XAS spectra measured. For this, a significant number of unoccupied states had to be added to the calculation. The total number of states, NBAND (occupied plus unoccupied), is also shown in Table~\ref{tab:conv_param_xas}.

Notice that VASP does not include spin-orbit coupling in the $2p$ core hole. Therefore, the $L_2$ ($j = 1/2$) and $L_3$ ($j = 3/2$) edges are represented by the same feature in the theory.

For the density of state (DOS) computation, a supercell is not necessary, but we used a denser $k$-point grid, as detailed in Table~\ref{tab:conv_param_xas}, to achieve a similar resolution in energy as in the XAS computation. The energy cutoff for the plane wave basis is the same as the one used for the simulation of the XAS, but the total number of states per $k$-point is significantly lower while the number of $k$-points is increased.

\section{Results}
\subsection{X-ray absorption spectroscopy}
\label{sec:xray}
The Ag $L_3$ absorption edge is due to transitions from the core $2p_{3/2}$ states to $4d$ or $5s$ unoccupied states. Typically, the transition to empty $d$ states leads to preedge peaks in X-ray absorption, commonly referred to as ``white lines" due to their appearance in old photographic detection. 

Fig.~\ref{fig:XANES} shows the XAS spectra from the present work for powder samples (full lines)  and from previous measurements by Miyamoto et al. (dot-dashed lines) \cite{Miyamoto2010}. The \ch{Ag2O} and AgF spectra are in good agreement with Ref.~\cite{Miyamoto2010}, although our measurements show sharper features and more detail, probably because of the different techniques used. Black lines are the theoretical computations in the SCH approximation.

Although the spin-orbit interaction, which splits the $L_2$ and $L_3$ edges by 172 eV is not considered in the calculations, the energy of the theoretical edge is surprisingly close to the experiment for the $L_3$ edge. A shift in the range $-2\sim 4$ eV was enough to align the theoretical data to the experiments.
Another important aspect neglected by the theory is lifetime effects, which will broaden narrow peaks in the theory. Despite those shortcomings, the theory reproduces several features of the data.  

\subsubsection{Ag}
\begin{figure}[tb]
\centering
\includegraphics[width=\columnwidth]{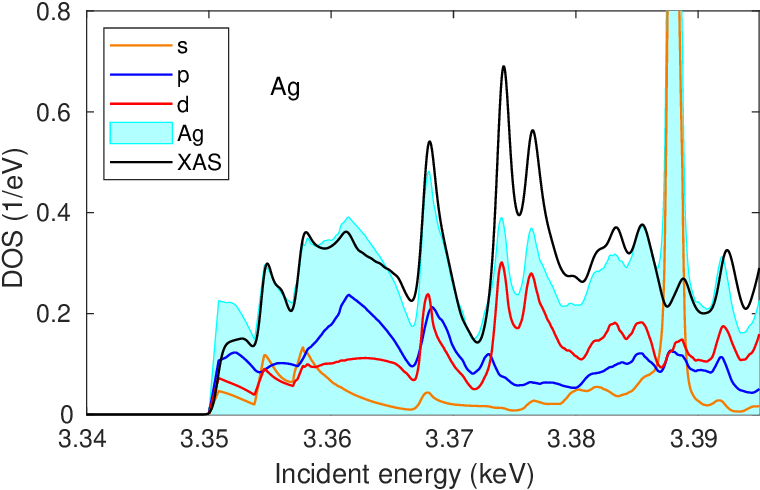}
\caption{\label{fig:dosAg}  Ag symmetry projected (orange, blue, red) DOS and total silver DOS (cyan shadowing) compared to the computation of the XAS in the SCH method shown in Fig.~\ref{fig:XANES} (black, intensity in arbitrary units). 
To simulate the edge, we show only the unoccupied part of the DOS, and the Fermi energy has been shifted to match the main features of the SCH computation.
}
\end{figure}

The experiment shows no white line consistent with the $4d^{10}5s$ configuration of pure Ag. The theory describes the overall edge shape quite well (Fig.~\ref{fig:XANES}) and exhibits some peaks at 3.368 keV, 3.374 keV, and 3.377 keV that correspond to broad structures observed in the experiment. Naively, one would expect the edge to have a dominant $5s$ character, but the DOS analysis (Fig.~\ref{fig:dosAg}) shows that there are significant $5p$ and $4d$ contributions due to strong hybridization. In general, the theoretical XAS spectra reflect main structures in the DOS except for the strong $s$ peak at high energy. 

\subsubsection{$4d^{10}$ compounds}
AgF data by Miyamoto {\it et al.} reproduced in Fig.~\ref{fig:XANES} (dot-dashed lines) show no evident pre-edge peak, although a weak structure at 3.352 keV was interpreted as a peak. 
We do observe a weak excitonic resonance at the edge of the absorption profile, which can be considered a precursor of the white line. We interpret this as due to the mixing of the nominally $d^{10}$ ground state with the $d^{9}s^1$ state mediated by the fluorine $p$ orbitals. Similar preedge peaks have been observed\cite{Grioni1992} in \ch{Cu2O} with a $3d^{10}$ ground state and in other Ag $4d^{10}$ compounds\cite{Miyamoto2010}. 

\begin{figure}[tb]
\centering
\includegraphics[width=\columnwidth]{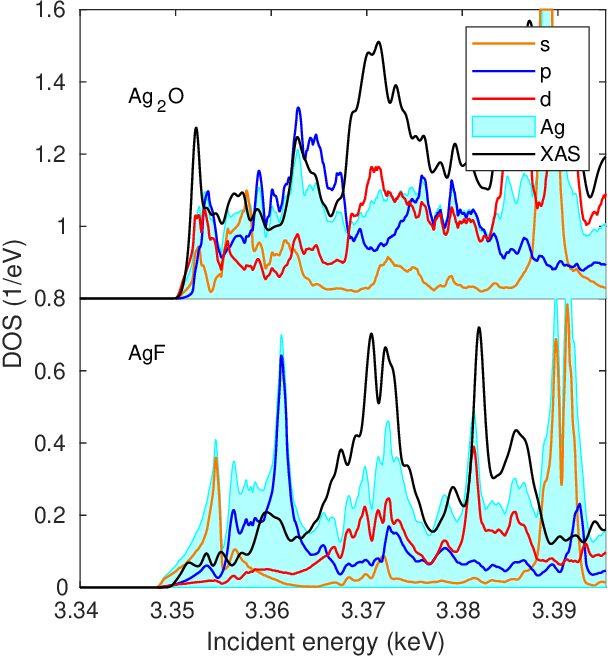}
\caption{\label{fig:dosd10} Same as Fig.~\ref{fig:dosAg} for $d^{10}$ compounds. The \ch{Ag2O} data has been shifted vertically by 0.8~eV$^{-1}$.
}
\end{figure}

The SCH computation reveals a gradual increase in intensity from the edge, accompanied by intense structures at high energies (3.36 keV, 3.37 keV, and 3.382 keV), which also corresponds to broad structures observed in the experiment. We attribute the very weak excitonic resonance to the small density of states at the edge, which has mainly $5s$ character (Fig.~\ref{fig:dosd10}). 
In contrast, the DOS in \ch{Ag2O} exhibits a steep rise at the edge region, which explains the strong excitonic resonance both in the theory and in the experiment. 
Also, in this case, the resonance is sharper than the results from Ref.~\cite{Miyamoto2010} (c.f. Fig.~\ref{fig:XANES}).

In \ch{Ag2O} the DOS shows a weaker $s$ contribution and more weight of other symmetries due to stronger hybridization. The difference between the AgF and \ch{Ag2O} DOS reflects the fact that ligand $2p$ orbitals are located at a higher energy with respect to Ag $4d$ orbitals, which is the main motivation to replace O by F mentioned in the introduction.

\begin{figure}[tb]
\centering
\includegraphics[width=\columnwidth]{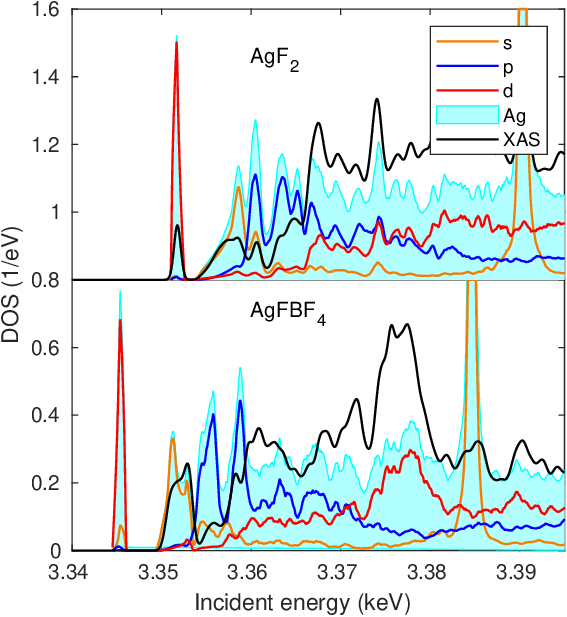}
\caption{\label{fig:dosd9} Same as Fig.~\ref{fig:dosAg} for $d^{9}$ compounds. The \ch{AgF2} data has been shifted vertically by 0.8~eV$^{-1}$.
}
\end{figure}

\subsubsection{$4d^{9}$ compounds}
{The experiment for \ch{AgF2} and \ch{AgFBF4} shows clear white lines at nearly the same energy (3.348 keV) consistent with a $d^9$ ground state.

For the theoretical computations, we considered an antiferromagnetic state. 
In this case, we obtain very weak white lines compared with experiments. Despite this deficiency, in the case of  \ch{AgF2} the white line and higher energy structures are located in reasonable agreement with the experimental features. 

The experiment in \ch{AgFBF4} 
shows two preedge features: Besides the narrow white line at 3.348 keV there is a broader feature at 3.352 keV. Aligning this with a similar structure in the theory yields to a strong misalignment of the weak white line of the theory with the strong white line of the experiment. Also for high energy features the agreement is poor.

A hint for the reason for the weak white lines in the theory is obtained by
plotting simply the unoccupied DOS without the core hole in a non-magnetic computation. This yields a result similar to the experiment  (Fig.~\ref{fig:dosd9}). Indeed, 
because $5d$ are narrow and only 1/10 of the $5d$ DOS is unoccupied, one obtains a narrow feature which mimics the experimental white line. As expected, this is almost purely $5d$-character. The black lines in Fig.~\ref{fig:dosd9} shows the results of the SCH in the non-magnetic case, showing again that this approximation yelds very weak white lines.

The significantly weaker white lines within the SCH computation is related to screening effects. Indeed, the core hole acts as a positively charged impurity. Relaxing the charge,  metallic electrons fill the ``impurity" state, which therefore becomes Pauli blocked as a possible final state, strongly reducing the intensity of the white line.

The screening is ineffective in the experiment because the core hole lifetime is very short, manifesting as a lifetime broadening. For the Ag $L_3$ core state, the FWHM\cite{Krause1979} is 2.4~eV. Particle-hole excitations with energy below this scale will not have time to relax and can be considered frozen. This explains why  electrons close to the Fermi level are not efficient at screening the exciton in the experiment. In the DOS computation, this screening is not present, explaining the appearance of an intense feature. 
A more accurate computation should consider core-hole lifetime and excitonic effects on an equal footing and is beyond our present scope. }

Because of the symmetry of the structure [Fig.~\ref{fig:AgF2_structure}(c)], the \ch{AgFBF4} theoretical spectra for the $x$ and $y$ polarizations are identical and different from the $z$ polarization. We find that the overall form of the absorption has a weak dependence on polarization (not shown) except for the broad feature at 3.352~KeV in Fig.~\ref{fig:XANES}, which is 10 times more intense for the electric field oriented perpendicular to the chain ($x,y$) than along the chain direction ($z$).

For \ch{AgF2}, the theoretical white line intensity shows some anisotropy for different polarizations (50\% changes in the intensity, not shown). A similar feature as the broad pre-edge feature of \ch{AgFBF4} is present in \ch{AgF2}  at 3.356 keV, showing also a nearly 10 times anisotropy in intensity but with the stronger feature when the electric field is along the $z$-direction ($c$-axis in Fig.~\ref{fig:AgF2_structure}). This inversion of the anisotropy is consistent with the nature of the ground-state hole found in DFT, which is 
$d_{3z^2-r^2}$ for \ch{AgFBF4} and $d_{x^2-y^2}$ for \ch{AgF2}. 
These anisotropic effects can be the subject of a future study in oriented single crystals. 

\begin{figure*}[t]
\centering
\begin{subfigure}[t]{0.02\textwidth}
    \textbf{(a) }
\end{subfigure}
\begin{subfigure}[t]{0.47\textwidth} \includegraphics[width=\linewidth,valign=t]{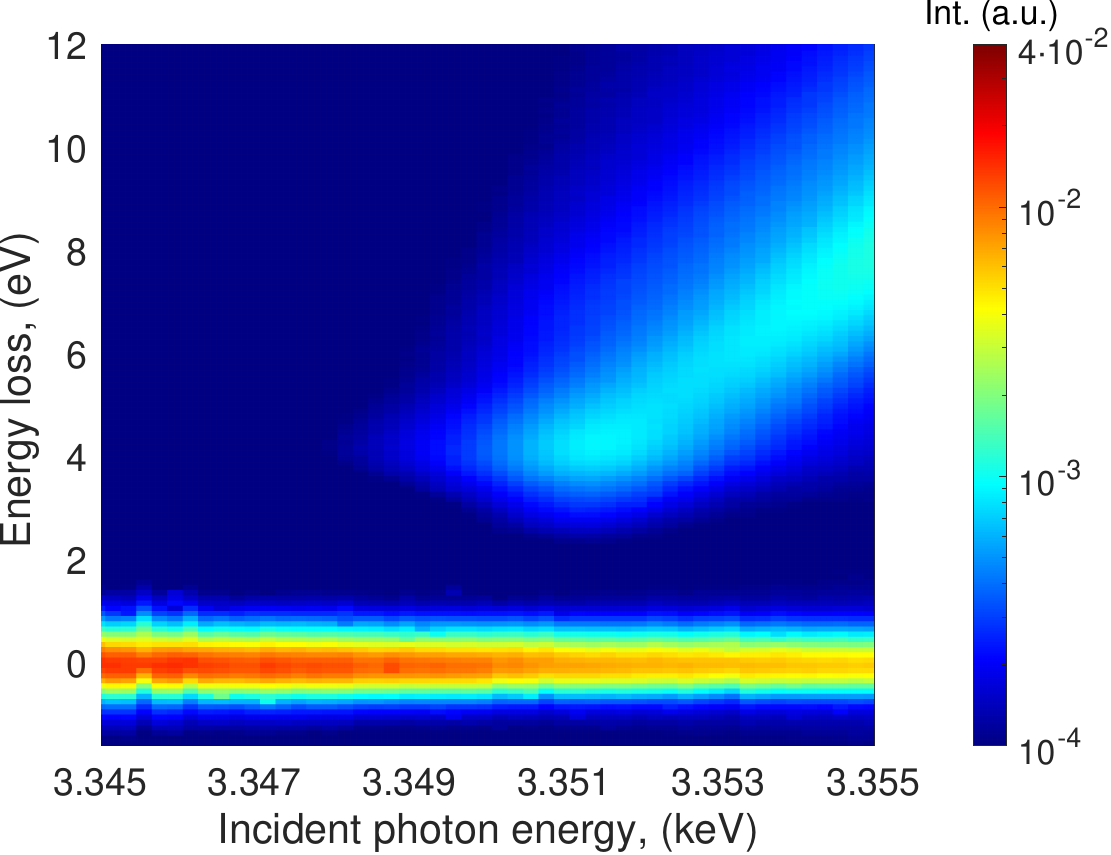}
\end{subfigure}\hfill
\begin{subfigure}[t]{0.02\textwidth}
    \textbf{(b) }  
\end{subfigure}
\begin{subfigure}[t]{0.45\textwidth}

\includegraphics[width=\linewidth,valign=t]{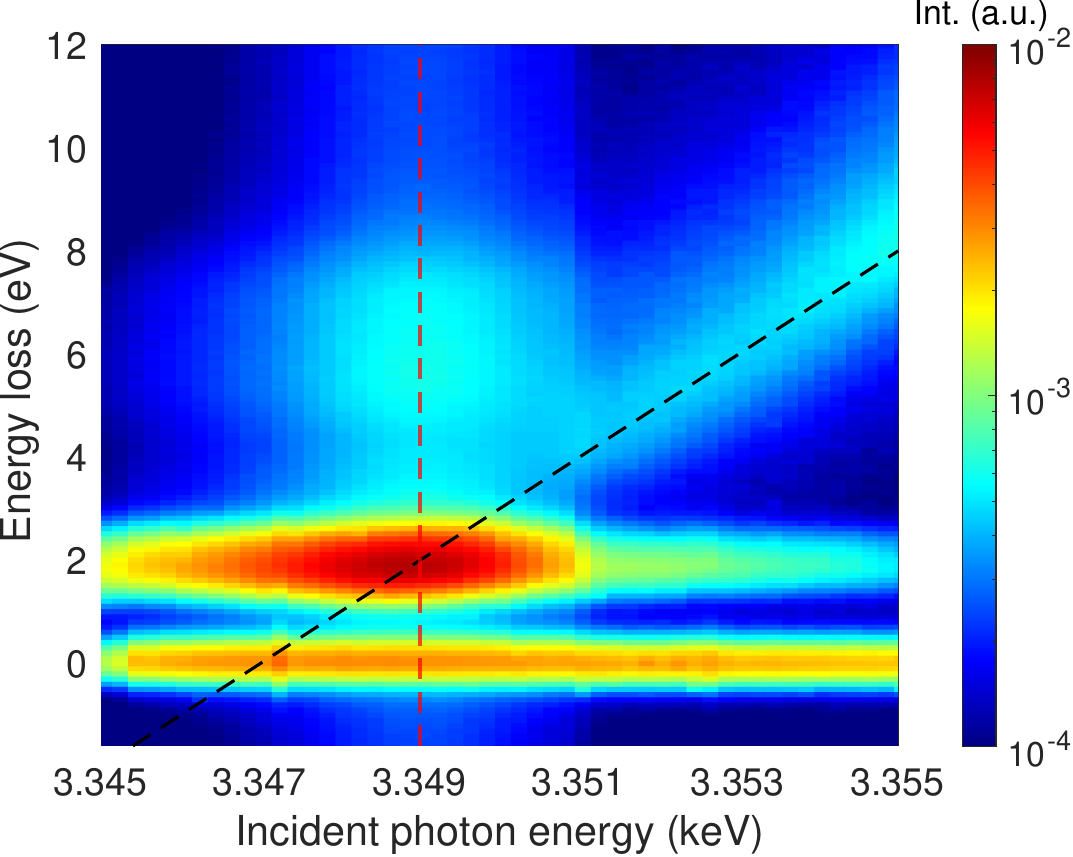}
\end{subfigure}
 \caption{RIXS/fluorescence intensity maps for  \ch{AgF} (a) and  \ch{AgF2} (b) powder samples in log-scale. %
The vertical grey dashed line shows the cut in Fig.~\ref{fig:AgF2_powder}. The oblique black-dashed line is the fluorescence scan used for the absorption measurement. Notice the overlap with the strong $dd$ RIXS excitation.} \label{fig:AgFAgF2map}
\end{figure*}

\subsection{Resonant inelastic X-ray scattering\label{sec:rixs}}

Figure~\ref{fig:AgFAgF2map} shows the false color map of the RIXS/fluorescence spectra of AgF and \ch{AgF2}.
The dispersive features correspond to fluorescence, while the non-dispersive ones are Raman contributions to the RIXS spectra.

\ch{AgF2} shows a strong feature at 2~eV of energy loss corresponding to $dd$ transition and a faint structure in the 4-8 energy loss range corresponding to charge transfer excitations (notice the log intensity scale). These assignments are based on previous computations and measurements at the fluorine $K$ edge~\cite{Bachar2022} as discussed in more detail below.  In contrast, for AgF, these features are not present; only the elastic line and the dispersive fluorescence feature are visible.  
This behavior is consistent with the closed-shell nature of AgF in contrast to the open $4d$ shell of \ch{AgF2}. \ch{Ag2O} exhibits a map similar to AgF consistent with the $d^{10}$ nature and is not shown. Likewise the \ch{AgFBF4} map (not shown) resembles  \ch{AgF2} reflecting the $d^{9}$ ground state.

{The black dashed line in panel (b) corresponds to the scan done at constant outgoing photon energy in the HERFD-XAS experiments. Such a scan intercepts a strong $dd$ process, which is the white-line identified in Fig.~\ref{fig:XANES}. 
Notice that, strictly speaking, HERFD-XAS is not a measurement of the absorption coefficient, but it can be shown to be closely related to it. This is related to the fact that photon absorption is the first step of the RIXS process. Therefore, the same feature can be interpreted in different ways depending on how the scan is done: a $dd$ excitation for a vertical scan in Fig.~\ref{fig:AgFAgF2map}, 
or an absorption white line for the diagonal scan (constant outgoing photon energy).   
}

For \ch{AgF2} in Fig.~\ref{fig:AgFAgF2map}(b), the maximum RIXS intensity is obtained for an incident photon energy of 3.349~keV. 
Therefore, in the following, we concentrate on the spectra at this incident energy.

\begin{figure}[tb]
\centering
\includegraphics[width=\linewidth,valign=t]{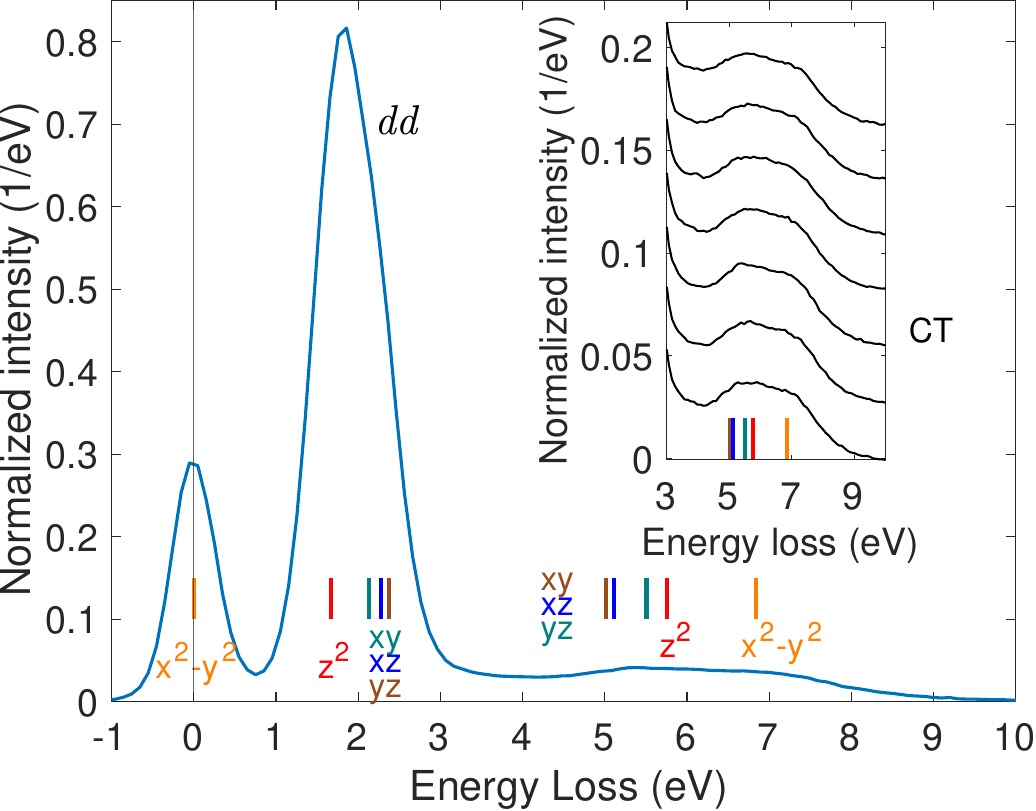}
 \caption{Ag $L_3$ RIXS spectrum of powder \ch{AgF2} sample (blue curve) - measured at 3.349 keV incident energy. Colored vertical lines represent bonding and antibonding states for hybridized orbitals between $d^9$ and $d^{10}L$ configurations following cluster computation done in \cite{Bachar2022} with the colors identifying the symmetry of the states ($z^2\equiv 3z^2-r^2$). 
The inset shows the CT part of the spectrum for different incoming photon energies. The lower curve is for an incoming photon energy of 3.3485 keV, and the following curves correspond to steps of 0.2 eV successively shifted by $1.5\times10^{-4}$ (normalized intensity, 1/eV). All the curves are normalized over $dd$+CT area. A linear background was subtracted.} \label{fig:AgF2_powder}
\end{figure}

In Fig.~\ref{fig:AgF2_powder}, we show the RIXS spectrum corresponding to a photon energy of 3.349 keV. The inset shows the CT part of the spectrum for different photon energies, confirming that it does not disperse and therefore it is not a fluorescence feature. The vertical lines in the main panel are the results of the exact diagonalization computations from Ref.~\cite{Bachar2022,Piombo2022} showing the position of one-hole states in a \ch{AgF6} cluster.  We write the one-hole states as, 
\begin{equation}\label{eq:defpsi}
\begin{split}
\ket{\psi_d(\nu)\sigma}&=\alpha_{\nu} \ket{d^9,\nu\sigma}+\beta_{\nu} \ket{d^{10}\underline{L},\nu\sigma}\\ 
\ket{\psi_p(\nu)\sigma}&=-\beta_{\nu}  \ket{d^9,\nu\sigma}+\alpha_{\nu} \ket{d^{10}\underline{L},{\nu}\sigma}
\end{split}
\end{equation}
where 
$\nu=x^2-y^2,3z^2-r^2,xy,yz,zx$ and $\sigma$ denotes the spin. The states on the right are the ionic configurations, where
$\ket{d^{10}\underline{L},{\nu}\sigma}$ denotes a $d^{10}$ state with a hole in a linear combination of ligand $p$ orbitals that transforms like one of the $d$ orbitals\cite{Eskes1990,Bachar2022}. 
$\alpha_{\nu}$, $\beta_{\nu}$ are real coefficients and without loss of generality we take $\alpha_\nu^2>\beta_\nu^2$ so $\ket{\psi_d(\nu)\sigma}$ ($\ket{\psi_p(\nu)\sigma}$) has prevalent $d^9$ ($d^{10} \underline{L}$) character.
Each pair ($\ket{\psi_d(\nu)\sigma}, \ket{\psi_p(\nu)\sigma}$) represents a bonding-antibonding combination for one hole states in $D_{4h}$ symmetry. This is not an exact symmetry in \ch{AgF2}, but it can be assumed within the \ch{AgF6} cluster to a good approximation.

The ground state has approximate $d_{x^2-y^2}$ symmetry and is represented by the vertical line at zero energy in Fig.~\ref{fig:AgF2_powder}. 
The excited states near 2~eV correspond to the states with the hole in orbitals of prevalent Ag $d$ character and different symmetry, as indicated by the labels. Transitions between the ground state and these states are classified as $dd$ transitions. The excited states between 4~eV and 6~eV have a prevalent hole ligand character, so transitions from the ground state to these states correspond to charge-transfer transitions. 
Comparison of the cluster computation with the spectrum confirms the peak at 2~eV as the $dd$ excitations and the broad feature at high energy as the CT part. 

\begin{figure}[tb]
        \centering        \includegraphics[width=\linewidth]{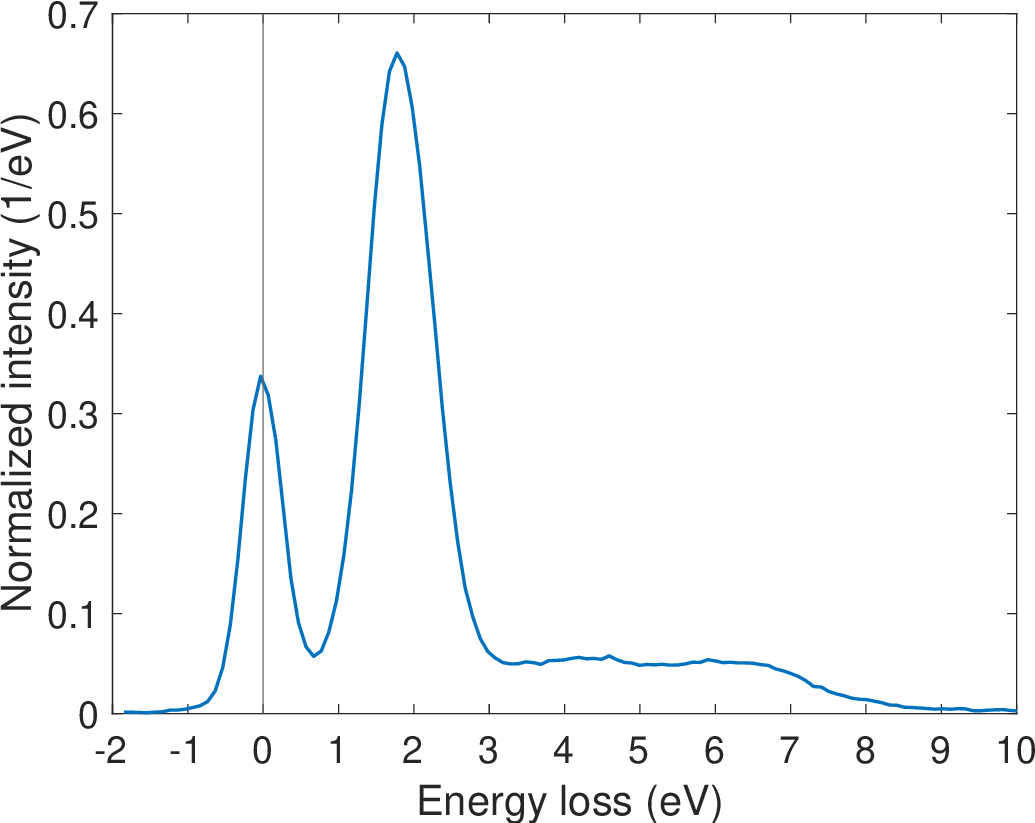}
    \caption{
RIXS spectrum for powder \ch{AgFBF4} sample measured with 3.349 keV incident photon energy. The curve was normalized over $dd$+CT area. A linear background was subtracted.}
\label{fig:AgFBF4_pow}
\end{figure}

Figure~\ref{fig:AgFBF4_pow} shows the RIXS spectrum for \ch{AgFBF4}. The spectrum shows similar $dd$ and CT features as \ch{AgF2}. One difference is that the CT band shows a weak feature at 4~eV energy loss, where the \ch{AgF2} spectrum shows a minimum (see also the inset of Fig.~\ref{fig:AgF2_LCO_all} below). 
Also in this case, we have verified that the CT feature exhibits no signs of dispersion (not shown), thereby excluding fluorescence contamination. A cluster computation has not been performed in this case; however, the experimental results indicate similar hybridizations and charge transfer energies as for \ch{AgF2} (c.f. Fig.~\ref{fig:AgF2_powder}). Due to the sensitivity of the samples to radiation, we cannot exclude the possibility that part of the boron was released as \ch{BF3} and the sample became partially \ch{AgF2}, thereby increasing its similarity to this compound.

\begin{figure*}[tb]
    \centering   \includegraphics[width=\textwidth]{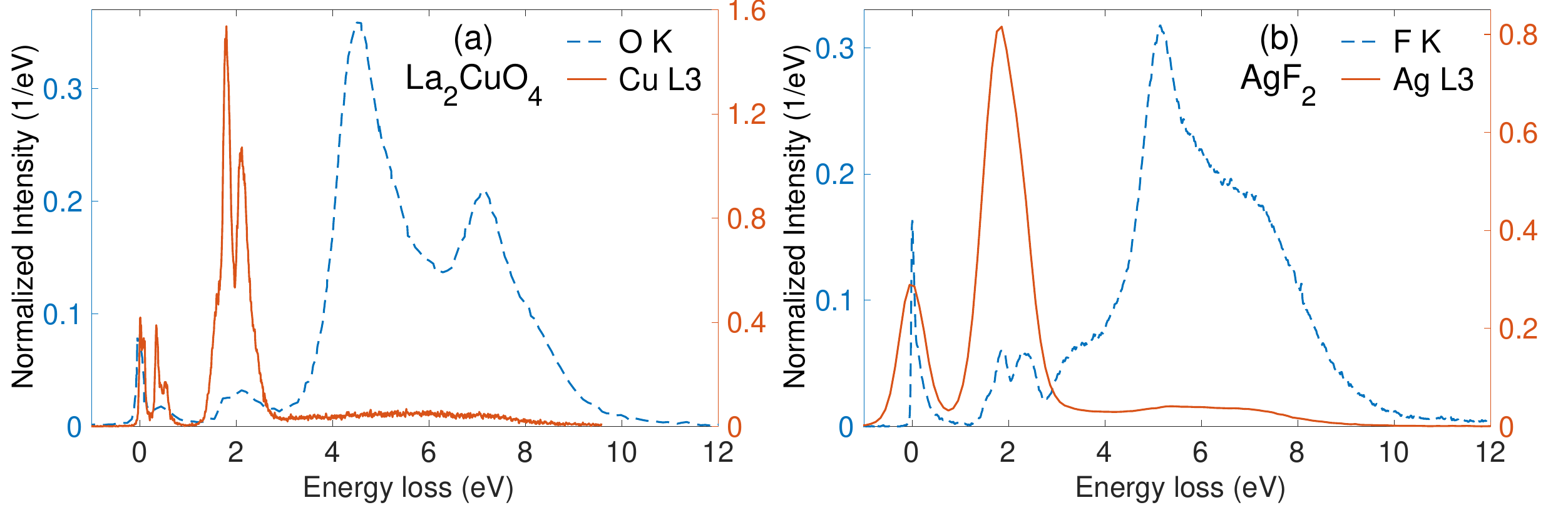}
    \caption{Comparison between RIXS spectra for \ch{La2CuO4} (a) and \ch{AgF2}  (b) measured at both edges (K,L). The \ch{AgF2} Ag $L_3$ data is the present measurement. The other curves are based on reference data from Refs.~\cite{Bachar2022} (\ch{AgF2} $K$ edge), \cite{Bisogni2012I} (LCO $K$ edge), and \cite{Martinelli2022fractional} (LCO L edge). All the curves are normalized over $dd$ + CT area.
    }
    \label{fig:L3_vs_K_all}
\end{figure*}
\subsection{Observability of spin-flip process}
It was suggested long ago that spin-flip processes may be excited in RIXS\cite{degroot1998} because of the strong spin-orbit interaction in the transition metal core-hole. It was later shown \cite{Ament2009} that for the spin-flip process to be allowed in $d^9$ systems with a $d_{x^2-y^2}$ hole state (e.g., parent cuprates and \ch{AgF2}), the magnetic moment must lie in the basal plane. This is the case in cuprates and also \ch{AgF2} according to early neutron scattering experiments\cite{Fischer1971II}.

The transitions that make the spin flip possible are the following: 
\begin{equation}
    \ket{x^2-y^2 \uparrow_x}\xrightarrow{xE_x}\ket{x \uparrow_x}\xrightarrow{S_zL_z }\ket{y \downarrow_x} \xrightarrow{yE_y} \ket{x^2-y^2 \downarrow_x}.
\end{equation}
Here $\ket{x^2-y^2 \uparrow_x}$ is a shorthand for the state $ \ket{d^9,x^2-y^2 \uparrow_x}$ defined above but
with the spin polarized in the $x$  direction, $xE_x$ represents an electric dipole transition with the photon electric field in the $x$ direction, $\ket{x \uparrow_x}$ represents the $d^{10}$ state with a core hole in a $p_x$ state, 
$S_z L_z $ is the spin-orbit coupling with the term $ S_z (x\partial_y-y\partial_x)$, which flips both the direction of the spin and the $p$ orbital. 
The remaining terms follow an analogous notation. The two dipole transitions correspond to the incoming and the outgoing photon electric field, and the chain of transitions illustrates the well-known fact that the two polarizations should be crossed for the spin-flip transition to be allowed\cite{Ament2009}. 

For the case of \ch{AgFBF4} an important difference according to DFT\cite{Kurzydowski2017Prediction,Kurzydowski2017Large,Koteras2022} is that the ground state has prevalently  $d_{3z^2-r^2}$ character with the $z$ axis oriented along the chain,  different from parent cuprates and also one-dimensional cuprates as \ch{Sr2CuO3}. Long-range order has not been reported, consistent with a spin liquid ground state. We can still expect antiferromagnetic quasi-long-range order to develop with the magnetic moment predominantly in one spatial direction. Adding spin-orbit coupling, we find that at the DFT level, the spin is in the $x-y$ plane. In this case, the chain of transitions that can lead to a spin-flip process reads
\begin{equation}
    \ket{3z^2-r^2\uparrow_x}\xrightarrow{zE_z}\ket{z \uparrow_x}\xrightarrow{S_yL_y }\ket{x \downarrow_x} \xrightarrow{xE_x} \ket{3z^2-r^2 \downarrow_x}.
\end{equation}
This shows that the process is allowed with one polarization in the chain direction and the other polarization perpendicular to it.

Unfortunately, the quasielastic line is too broad in both compounds to disentangle these excitations. Still, the present considerations set the stage for future studies of magnetic excitations using higher-resolution measurements.

\subsection{\label{sec:Comparison} The electronic structure of \ch{AgF2} vs. \ch{La2CuO4}.}

Having available data in the Cu $L_3$ and O $K$ edge in LCO \cite{Bisogni2012I, MorettiSala2011}, and analogous data for the Ag $L_3$ (this work) and the F $K$ edge \cite{Bachar2022} we can now make a detailed comparison of the electronic structure of the two compounds. 

Fig.~\ref{fig:L3_vs_K_all}  shows the RIXS spectra in the metal and the ligand edge for LCO(a) and for \ch{AgF2}(b). Despite the much lower resolution of the present Ag $L_3$ data, the similarity of the spectra in the two materials is striking.

\begin{figure}[b]
    \centering   \includegraphics[width=\linewidth]{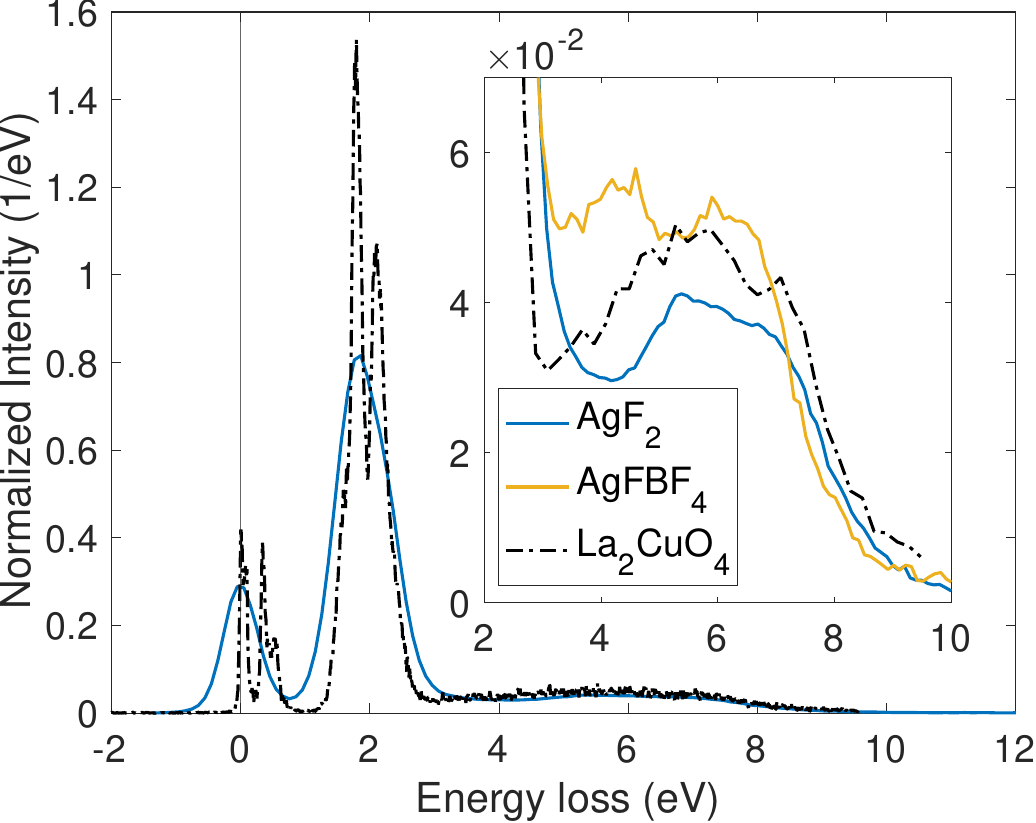}
    \caption{RIXS Ag $L_3$ spectrum for \ch{AgF2} powder sample (full line) compared with Cu $L_3$ spectrum of LCO (dotted-dashed line). The inset shows a zoom of the CT part for \ch{AgF2} and \ch{AgFBF4} compared with LCO. All the curves are normalized over $dd$+CT area. The curve for LCO shown in the inset was smoothed with a step of 0.2 eV. }
    \label{fig:AgF2_LCO_all}
\end{figure}

Figure~\ref{fig:AgF2_LCO_all} shows a comparison of the data in the $L_3$ edge of both compounds normalized to the CT+$dd$ total intensity. The $dd$-excitation appears much broader in \ch{AgF2} due to the poorer energy resolution, while the inset shows the surprising similarity of the CT band in \ch{AgF2}, \ch{AgFBF4}, and LCO. This clearly indicates that electronic parameters in these materials are very similar, apart from the following caveat. The main panel shows that the quasielastic peak of \ch{AgF2} likely contains magnetic and phonon excitations, which are visible in high-resolution experiments in LCO but cannot be separated in the present low-resolution experiments. From this comparison, it is clear that the zero of energy determined for \ch{AgF2},  in reality, corresponds to an average between magnetic excitations and the truly elastic line. Therefore, all Ag $L_3$ spectra should be shifted by some systematic error of the order of the difference between a broadened version of the cuprate data and our elastic peak $\sim0.3$~eV. 
In other words, we estimate that all excitation energy reported for the Ag $L_3$ edge should be increased by about that amount. However, since the relative weight between elastic and inelastic signals is unknown, and there is no systematic way to estimate this effect, we do not attempt such a correction. 
Notice that the excitations at the F $K$ edge measured at high resolution and shown in Fig.~\ref{fig:L3_vs_K_all}(b) appear systematically at higher energy with respect to the analogous features in the Ag $L_3$ edge, which we attribute to this effect. 

\subsection{Measuring covalency from intensity ratios}\label{sec:covalency}

Here, we present a method for evaluating the covalency of compounds based on the intensity ratios of their main features. To quantify the relative intensities of different features, we fitted all the spectra with Gaussians as shown in the App.~\ref{app:gaussians} (Fig.~\ref{fig:fitting_L3}). %

Relative spectral weights of the $dd$ and CT excitations are reported in Table~\ref{tab:dd_CT_int}. The quasielastic spectral weight was found to be too sensitive to the experimental conditions and is not included in the analysis.   Computed values of $dd$/CT intensity ratios measured at $L_3$ edges are similar between Ag compounds and reference LCO data.

\begin{table}[tb]
\begin{ruledtabular}
  \caption{Comparison of RIXS integral intensities for $dd$ and CT peaks for powder and single crystal samples from the present work and integrating previous spectra for the F $K$ edge data of \ch{AgF2}\cite{Bachar2022} and Cu $L_3$ and O K RIXS of LCO\cite{MorettiSala2011}. Integrals were computed by fitting the different features with Gaussians as shown in Fig.~\ref{fig:fitting_L3}. This yields approximately 4\% of total weight assigned to the CT band, which overlaps with the $dd$ excitations. Changing the criteria for assignment (for example, with a sharp cutoff at the minimum intensity) introduces an indeterminacy in the weights of this order.  All angles are in radians.    
    \label{tab:dd_CT_int}}
    \centering
    \begin{tabular}{c|cccc}
         Edge & Compound&Notes & $dd$ & CT  \\
    \hline
         \multirow{5}{*}{$L_3$}  & \ch{AgF2}& $\Delta \theta=\pi/2$& 79\%  & 21\%   \\
         & \ch{AgFBF4} &$\Delta \theta=\pi/2$ & 75\%   & 25\%  \\
         & \ch{La2CuO4} &$\theta'=0.68$ Ref.~\cite{Martinelli2022fractional}& 73\% & 27\% \\
    \hline
         \multirow{2}{*}{K} & \ch{AgF2} &Ref.~\cite{Bachar2022} & 9\% & 91\% \\
         & \ch{La2CuO4} &Ref.~\cite{Bisogni2012I} & 3\%  & 97\% \\
    \end{tabular}
    \end{ruledtabular}
\end{table}

The physical content of the relative spectral weights can be understood as follows. The RIXS cross-section is given by the Kramers-Heisenberg equation\cite{Kramers1925,Ament2011}. The Ag $L_3$ core state is quite broad (FWHM=2.4~eV\cite{Krause1979}), which implies a quite short lifetime. To leading order in the ultra-short hole lifetime approximation\cite{Luo1993,Ament2007,Bisogni2012I,Bisogni2012II} we can neglect the state dependence of the energy denominators. Then the total spectral weight of $dd$ excitations (including elastic and spin flip) is proportional to the sum of the matrix elements of  Ref.~\cite{MorettiSala2011},
\begin{equation}\label{eq:idd}
\begin{split}&I(dd)=\\    &\sum_{\nu\sigma\alpha=\pi,\sigma}|\bra{\psi_d(x^2-y^2)\downarrow}(\hat{\bm\epsilon}.\hat{\bm r}) P_{3/2}(\hat{\bm\epsilon}_\alpha'.\hat{\bm r}) \ket{\psi_d(\nu)\sigma}|^2
    \end{split}
\end{equation}
with the projector onto the core hole $j=3/2$ manifold $$P_{3/2}\equiv\sum_m \ket{d^{10}\underline{p_{3/2,m}}}\bra{d^{10}\underline{p_{3/2,m}}}, $$ and the dipole operator for incoming and outgoing photons written in terms of their polarization vectors, $\bm\epsilon$, $\bm\epsilon_\alpha'$ which we are assuming to be real vectors. The sum over $\alpha$ is over the possible polarizations of outgoing photons, as they are not discriminated in the experiment.

\begin{figure}
    \centering   \includegraphics[width=1\linewidth]{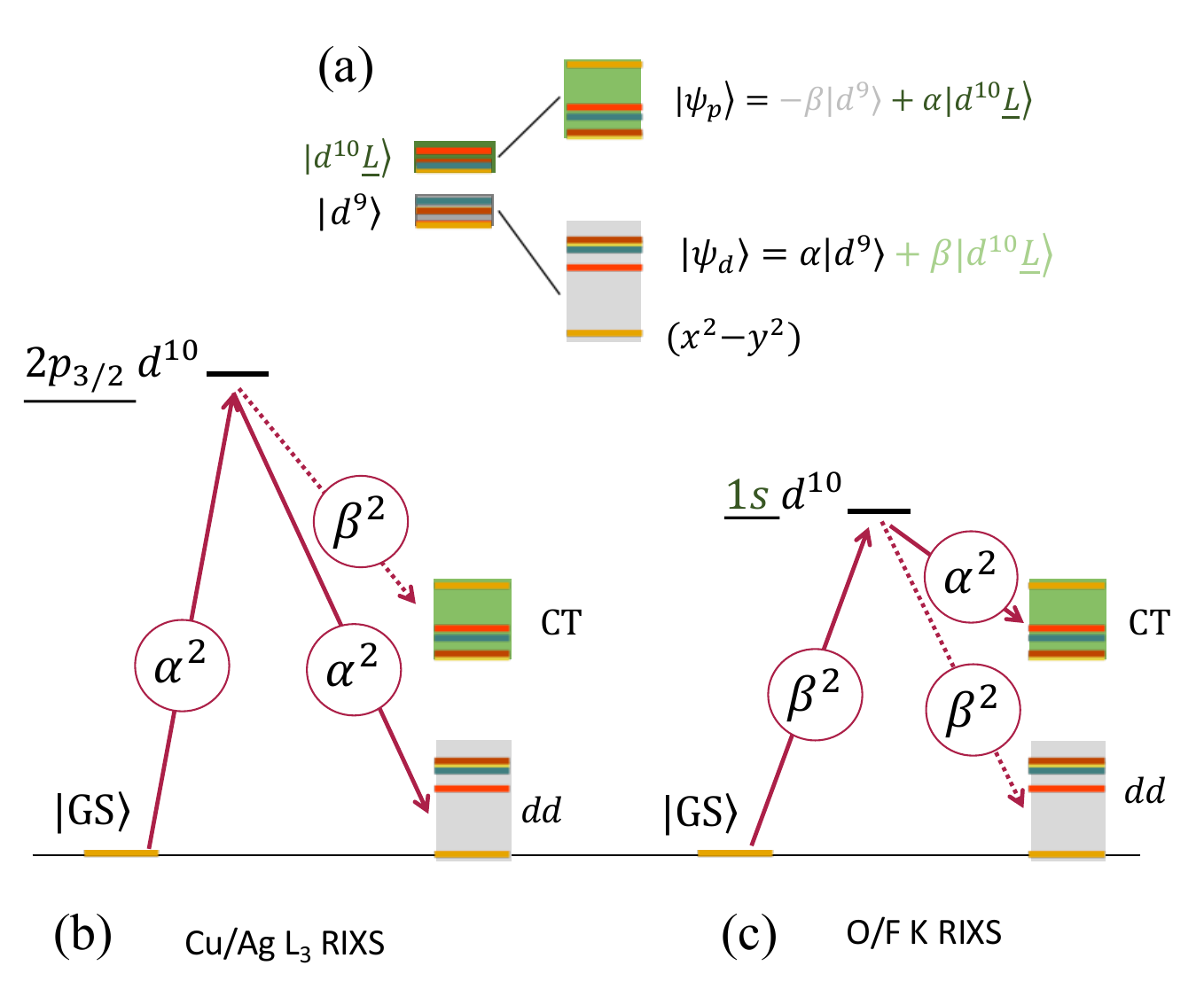}
    \caption{(a) Ionic levels (left) and hybridized levels (right) according to Eq.~\eqref{eq:defpsi}. The parameters to obtain these levels are the same as in Ref.~\cite{Bachar2022}. The hybridized levels are the same as shown in Fig.~\ref{fig:AgF2_powder}. (b) Processes involved for the transition metal $L_3$-edge RIXS. Circles indicate the probabilities of the transitions due to projection onto the ionic configurations. (c) Same for ligand $K$-edge RIXS. An underline in labels indicates a hole.}
\label{fig:RIXStransitions}
\end{figure}

Because the core hole is on the transition metal, the operators in Eq.~\eqref{eq:idd} act only on the $d^9$ component of the wave functions Eq.~\eqref{eq:defpsi}. Figure~\ref{fig:RIXStransitions}(a) shows the energy levels before (left) and after hybridization (right). 

Figure~\ref{fig:RIXStransitions}(b)  illustrates the RIXS process. The $\ket{d^9,x^2-y^2\sigma}$ component of the ground-state wave-function gets excited by the incoming photon with probability $\alpha_{x^2-y^2}^2$
to an intermediate $\ket{d^{10}\underline{p_{3/2},m}}$ state. Then, it emits a photon and gets deexcited to a $\ket{d^9,\nu\sigma}$ with probability $\alpha_\nu^2$ ($dd$ excitation), or to a $\ket{d^{10}\underline L,\nu\sigma}$ with probability $\beta_\nu^2$ (CT excitation).

The intensity can be put as, 
\begin{equation}\label{eq:idd2}
    I(dd)=\alpha_{x^2-y^2}^2\sum_{\nu}W_\nu \alpha_\nu^2
\end{equation}
with the weights computed with the ionic configurations, 
\begin{equation}
W_\nu=    \sum_{\sigma\alpha}
|\bra{d^9,x^2-y^2\downarrow}(\hat{\bm\epsilon}.\hat{\bm r}) P_{3/2}(\hat{\bm\epsilon}_\alpha'.\hat{\bm r}) \ket{d^9,\nu\sigma}|^2.
\end{equation}
Similarly, for the CT intensity, substituting $d\rightarrow p$ in the last ket in Eq.~\eqref{eq:idd} we obtain,
\begin{equation}\label{eq:iddfinal}
I(CT)=\alpha_{x^2-y^2}^2\sum_{\nu}W_\nu \beta_\nu^2
\end{equation}
Defining
\begin{equation}\label{eq:alpha2beta2}
\begin{split}
\overline{\alpha^2}\equiv&\sum_{\nu}w_\nu \alpha_\nu^2,\\
\overline{\beta^2}\equiv&\sum_{\nu}w_\nu \beta_\nu^2,\\
w_\nu\equiv&\frac{W_\nu}{W_T},\\
W_T\equiv&\sum_{\nu}W_\nu,
\end{split}
\end{equation}
we find 
\begin{equation}\label{eq:IdItot}
\frac{I(CT)}{I(dd)}=\frac{\overline{\beta^2}}{\overline{\alpha^2}}
\end{equation}
Since $\overline{\alpha^2}+\overline{\beta^2}=1$, we can estimate $\overline{\alpha^2}$, $\overline{\beta^2}$.
Because of technical reasons,  $I'(dd)$ was determined, excluding the quasielastic part. We can take this into account by correcting the intensity by the corresponding factor: 
\begin{equation}\label{eq:IdI}
\frac{I(CT)}{I'(dd)}=\frac{\overline{\beta^2}}{\overline{\alpha^2}}\left(\frac{1}{1-w_{x^2-y^2}\alpha_{x^2-y^2}/{\overline{\alpha^2}}}\right).
\end{equation}
The left-hand side is accessible experimentally, while the right-hand side can be computed for different models and different scattering geometries.

\begin{figure}[tbh]
    \centering   \includegraphics[width=.95\linewidth]{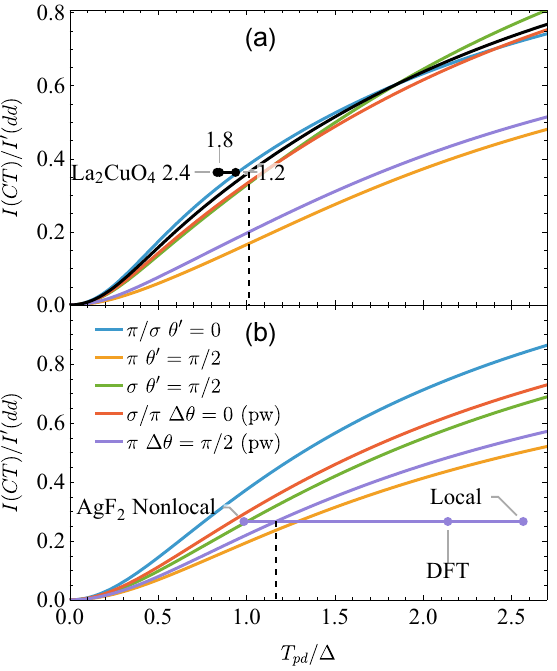}
\caption{CT to $dd$ RIXS intensity ratio vs. $T_{pd}/\Delta$ computed from Eqs.~\eqref{eq:alpha2beta2},\eqref{eq:IdI}  We show the results relevant for oriented crystal in $\pi/\sigma$ polarization for $\theta'=0$ (light blue) and $\pi$  and $\sigma$ incoming polarization for  $\theta'=\pi/2$ (orange and green, respectively). Averages for powder samples (pw) are for $\pi/\sigma$ incoming polarization with $\Delta \theta=0$ (red) and $\pi$ polarization for  $\Delta \theta=0$ (violet). Notice that powder results for $\sigma$ polarization are angle independent.  (a) Curves computed with LCO parameters with $T_{pp}/T_{pd}=0.54$\cite{Eskes1990}. 
The horizontal segment is the experimental ratio for LCO in $\sigma$ incoming polarization and an outgoing angle of $\theta'=0.68$ (rad). We also show the curve for the same outgoing angle (black).
The points indicate different theoretical values of $T_{pd}/\Delta$ labeled by the fundamental gap, $E_{\text{gap}}$ in eV from Ref.~\cite{Eskes1990}.  (b) Curves computed with \ch{AgF2} parameters from Ref.~\cite{Bachar2022}. 
 The horizontal violet line is the ratio for \ch{AgF2} from the present work. We indicate the value of $T_{pd}/\Delta$ estimated in Ref.~\cite{Bachar2022} using DFT and a Wannier analysis (DFT). The points marked ``Local" and ``Nonlocal" are alternative values of $\Delta/T_{pd}$ from  Ref.~\cite{Bachar2022}.  The vertical dashed lines in (a) and (b) indicate the deduced values of $T_{pd}/\Delta$ from the present analysis. See text for details. }
    \label{fig:ratiolasco}
\end{figure}

In Appendix~\ref{app:weights} we present a computation of the weights for general geometries and averaged for powder samples. For oriented crystals, we take the  $x,z$ plane to be the scattering plane with the incoming ($\hat{\bm k}$) and outgoing 
 ($\hat{\bm k'}$)  versors  forming an angle $\theta$ and $\theta'$ respectively with the $z$ axis. i.e. 
$\hat{\bm k}=(\sin\theta,0,\cos\theta)$,
$\hat{\bm k}'=(\sin\theta',0,\cos\theta')$. We define $\Delta\theta\equiv\theta-\theta'$.
Table~\ref{tab:rweights} shows the relative weights $w_\nu$ for specific geometries and $\pi$, $\sigma$ incoming polarization. Normal ($\theta'=0$) and grazing ($\theta'=\pi/2$) emission provide extreme values of the general weights presented in  Table~\ref{tab:weights}. 

\begin{table}
\caption{\label{tab:rweights} Relative weights $w_\nu$ for different scattering geometries and initial polarizations (pol) appropriate for oriented crystals, and averages for powder samples. Overlines are a reminder that a powder average has been performed.} 
$$
\begin{array}{c|ccccc}
\hline\hline
          &  \multicolumn{3}{c}{\text{Crystal}}      &     \multicolumn{2}{c}{\text{Powder}}      \\    
(\text{pol},\theta')& (\pi/\sigma,0)& (\pi,\pi/2)&(\sigma,\pi/2)&  &\\    
(\text{pol},\Delta\theta)& & & &(\pi/\sigma\footnote{Valid for any angle in $\sigma$ polarization.},\overline0)&(\pi,\overline{\pi/2}) \\ 
\hline
 x^2-y^2   & 0.366 & 0.079 & 0.293 & 0.307 & 0.136 \\
 3z^2-r^2  & 0.122 & 0.132 & 0.195 & 0.147 & 0.156 \\
{xy}       & 0.366 & 0.316 & 0.073 & 0.217 & 0.317 \\
{xz}       & 0.073 & 0.158 & 0.366 & 0.165 & 0.196 \\
{yz}       & 0.073 & 0.316 & 0.073 & 0.165 & 0.196 \\
\hline\hline
\end{array}
$$
\end{table}

The more important parameter determining the covalency of the material is the ratio of the hybridization in the $x^2-y^2$ sector ($T_{pd}$) to the charge transfer energy ($\Delta$).
The remaining $p-d$ hybridizations are assumed to be proportional to $T_{pd}$. 
Figure~\ref{fig:ratiolasco} shows the intensity ratio for different geometries as a function of $T_{pd}/\Delta$ for cuprates (a) and \ch{AgF2} (b). Other parameters needed for the computation are the crystal field splittings of the ionic $d^9$ configurations and the splittings of the symmetrized $d^{10}\underline{L}$ configurations due mainly to the $p-p$ hybridization. 
In Ref.~\cite{Eskes1990} for cuprates,  the first is neglected and the latter is controlled by the parameter $T_{pp}$. In the plot, we varied $T_{pp}$ in the same proportion as $T_{pd}$, but our definition of $\Delta$ as in Ref.~\cite{Bachar2022} is given by the difference of the averaged multiplet energies. This introduces a small contribution from $T_{pp}$, which we keep fixed. In analogy, for the case of \ch{AgF2}, we keep the average distance between the two multiplets, $\Delta$, fixed and vary the splitting of the $d^{10}\underline L$ orbitals and the $T_{pd}$ term in the same proportion, while keeping the crystal field splitting of the $d^9$ orbitals fixed. For details of the model and parameters, see App.~\ref{app:param}.

We can now use the experimental ratios to determine the degree of covalency of the compounds parametrized by the ratio $T_{pd}/\Delta$.
In the case of the cuprate, we need the curve for the specific geometry of the experiment in an oriented crystal [black line in Fig.~\ref{fig:ratiolasco}(a)]. The horizontal black segment indicates the experimental ratio, so the intersection with the curve gives the estimated value of $T_{pd}/\Delta$.
The obtained value is close to the values deduced decades ago using high-energy spectroscopies and cluster computations, as indicated by the labeled black points. 
The labels are the fundamental gap, $E_{\text{gap}}$  of Ref.~\cite{Eskes1990}. Three different parameter sets were proposed in this work, and experimental and theoretical results are closer for the parameter set with the smaller $E_{\text{gap}}$. 
If one tries to fit the position of the structures, the parameter with the higher 
$E_{\text{gap}}$ is the one which fits best (see Ref.~\cite{Bachar2022}).
However, the tree parameter sets are close, and the vertical distance between the segment and the curve is of the order of the error due to different criteria to separate CT and $dd$ excitations. Here, the separation is performed with the Gaussian fit of Fig.~\ref{fig:fitting_L3}. 
Given the uncertainties involved, we consider the agreement very good, which validates this method to obtain the covalency parameter $T_{pd}/\Delta$. 

Having established the usefulness of the analysis, we now switch to the case of \ch{AgF2} shown in Fig.~\ref{fig:ratiolasco}(b). The horizontal violet line is the intensity ratio for the present experiment, which should intersect with the violet line. The vertical black dashed line indicates the deduced value of $T_{pd}/\Delta$.

In this case, two different parameter sets were proposed from the analysis of F $K$ edge RIXS, optical spectroscopy\cite{Bachar2022}, and high-energy spectroscopy experiments\cite{Piombo2022}, while a third parameter set can be deduced taking bare DFT values from the Wannier analysis of Ref.~\cite{Bachar2022} and labeled as DFT in Fig.~\ref{fig:ratiolasco}(b).

The DFT computations of  Ref.~\cite{Bachar2022} suggested a very covalent scenario. In order to fit the position of the RIXS features, the covalency was increased even more, leading to the ``local" parameter set showing a small value of $E_{gap}$. Instead, the gap in the optical conductivity suggested a more ionic picture for unbound particle-hole excitations, leading to the ``nonlocal" parameter set. Both scenarios were made compatible by invoking exciton effects, which lower the gap for local particle-hole excitations. The covalency parameter for all three parameter sets is indicated in Fig.~\ref{fig:ratiolasco}(b) by the violet points. Intersection with the violet line (powder average) yields the estimated value of the covalency parameter for the present experiment. 

The present analysis points to an intermediate value of $T_{pd}/\Delta$ but closer to the nonlocal parameter set. This indicates a very similar, but somewhat higher degree of covalency in \ch{AgF2} than in LCO.  Notice that the different geometry of the experiments compensates for the higher intensity ratio of the latter.

Qualitatively, in a similar analysis for the $K$ edge of the ligand, one expects the role of $\alpha$ and $\beta$ to be interchanged as shown schematically in Fig.~\ref{fig:RIXStransitions}. In addition, one should consider the non-bonding orbitals. Furthermore, in this case, the dispersion of $p$ states becomes important, and modeling beyond a cluster model becomes relevant, which is outside our present scope. In any case,  the larger relative weight of the $dd$ transitions in the $K$ edge of \ch{AgF2} with respect to LCO (Table~\ref{tab:dd_CT_int} suggests that \ch{AgF2} is the more covalent material. However, as discussed in Ref.~\cite{Bachar2022}, in this case, also Coulomb intersite matrix elements can influence the intensity.

\FloatBarrier
\section{\label{sec:conc} Conclusions}

We have presented XAS and RIXS measurements on the Ag $L_3$ edge of three silver fluorides (AgF, \ch{AgFBF4}, \ch{AgF2}) and \ch{Ag2O} as a reference. The XAS results were compared with DFT computations, which allowed us to identify the origin of the main features in the spectra. 
The approximation used with a localized frozen core hole worked relatively well for closed-shell systems {but strongly underestimated the spectral weight of white lines in $d^9$ systems. We attributed this problem to the deficient treatment of core hole lifetime effects and an overestimation of the screening of the core hole by the conduction electrons, which calls for more accurate treatments.  }

The \ch{AgF2} $L_3$ RIXS results complement previous $K$ edge RIXS measurements. In addition, we present the RIXS spectrum of \ch{AgFBF4}, which is believed to host a spin-liquid ground state. 

The Ag $L_3$ edge belongs to the tender X-ray range (3.35 keV), making it a challenging experiment in commonly available facilities. In addition, the strong reactivity of silver fluorides and the sensitivity to radiation damage further complicate the measurements. Despite these difficulties, very informative RIXS spectra were obtained. 

The $K$ edge measurements of Ref.~\cite{Bachar2022} already demonstrated a striking similarity between the cuprate and the fluoride electronic structure.  The present measurements independently confirm this trend.

Notwithstanding the low resolution available in the present $L_3$ measurements, the silver fluorides exhibit clear $dd$, and CT features very similar to those in cuprates. The positions of peaks coincided very well with the results of previous cluster computation and the measurements at the $K$ edge\cite{Bachar2022}. 

The results were extended to the very interesting one-dimensional quantum antiferromagnet  \ch{AgFBF4}. The positions of the $dd$ and CT peaks were located at very similar energies as for \ch{AgF2}, suggesting that the octahedral-like geometry primarily determines the local electronic structure and depends weakly on the global structural details. Unfortunately, a partial sample degradation into \ch{AgF2} could not be excluded.
High-resolution studies would be most welcome here because a better characterization of the spectral differences in the spectra would enable verification of the sample integrity.   

We argue that the spectral weight ratio between the CT and $dd$ features is a measure of the covalency of the materials. 
We presented a method to systematically evaluate the degree of covalency for arbitrary geometries in $d^9$ compounds. 
The method was validated in cuprates, where parameters are well established, and then applied to silver fluorides. We also presented general formulae to perform the powder average of RIXS spectral intensities for arbitrary scattering matrices and linear polarization.

It is interesting that, for a negative charge transfer material, the ratio of intensities inverts\cite{Agrestini2024}, confirming this ratio as a tool for evaluating covalency. 

 The degree of covalency of cuprates and silver fluorides appears very similar, with silver fluorides somewhat more covalent than cuprates. One should be cautious that covalency is evaluated in average, summing over all $dd$ and CT excitations. It does not necessarily imply that the covalency of the ground state $x^2-y^2$ sector is larger in the silver fluorides than in cuprates.  

For cuprates, the parameter $T_{pd}/\Delta$ estimated here from the RIXS intensity ratios is very similar to well-established values in the literature\cite{Eskes1990}. For both \ch{AgF2} and LCO, however, it is difficult to fit the peak positions and the intensity ratios with the same parameter set. This may be a problem with the cluster model. Indeed, a similar difficulty was encountered in fitting optical and RIXS experiments in Ref.~\cite{Bachar2022}, which was attributed to excitonic effects. 
It would be interesting to extend the present model to take into account many-body correlations.  

Our results also call for a systematic study of spectral weights in RIXS and their dependence on different resonant conditions.

The finding that RIXS experiments are feasible at the Ag $L_3$ edge of silver fluorides calls for high-resolution spectra enabling the investigation of low-energy excitations as magnons, phonons, and plasmons. Recently available single crystals\cite{Poczynski2019} may allow the characterization of the full momentum dependence of the excitations. 

In general, magnetic interactions are expected to increase with covalency. Assuming a magnetic mechanism in a hypothetical metallic silver fluoride should lead to a larger $T_c$ than in cuprates of similar structure, as suggested in theoretical studies\cite{Grzelak2020}. 
Overall, the strong similarity of the spectra in the two edges strongly supports the idea that, from the electronic structure point of view, \ch{AgF2} is an excellent analog of cuprates. 
This provides a strong motivation to pursue the synthesis of doped compounds in the search for new avenues for high-$T_c$ superconductivity and quantum magnetism.

\begin{acknowledgments}
J.L. is in debt with Jeroen Van Den Brink and 
Maryia Zinouyeva for enlightening discussions. Work supported by the Italian Ministry of University and Research through the project Quantum Transition-metal FLUOrides (QT-FLUO) PRIN 20207ZXT4Z and under the Grant of Excellence Departments to the Department of
Science, Roma Tre University.
Z. M. gratefully acknowledges the financial support of the Slovenian Research and Innovation Agency (research core funding No. P1-0045; Inorganic Chemistry and Technology).
 We acknowledge the CINECA award under the ISCRA initiative for the availability of high-performance computing resources and support through class C projects ABRID (HP10CKR6YT) and Frage (HP10CDJZ0O), as well as class B project FluSCA (HP10BAWWVS).
Polish authors appreciate support from the National Science Center (NCN), project OPUS M(Ag)NET (2024/53/B/ST5/00631).
\end{acknowledgments}

\appendix

\section{Fits of the RIXS spectra}\label{app:gaussians}
\begin{figure}[t]
    \centering   
    \begin{subfigure}[t]{0.48\linewidth}
        \centering        \includegraphics[width=\linewidth]
        {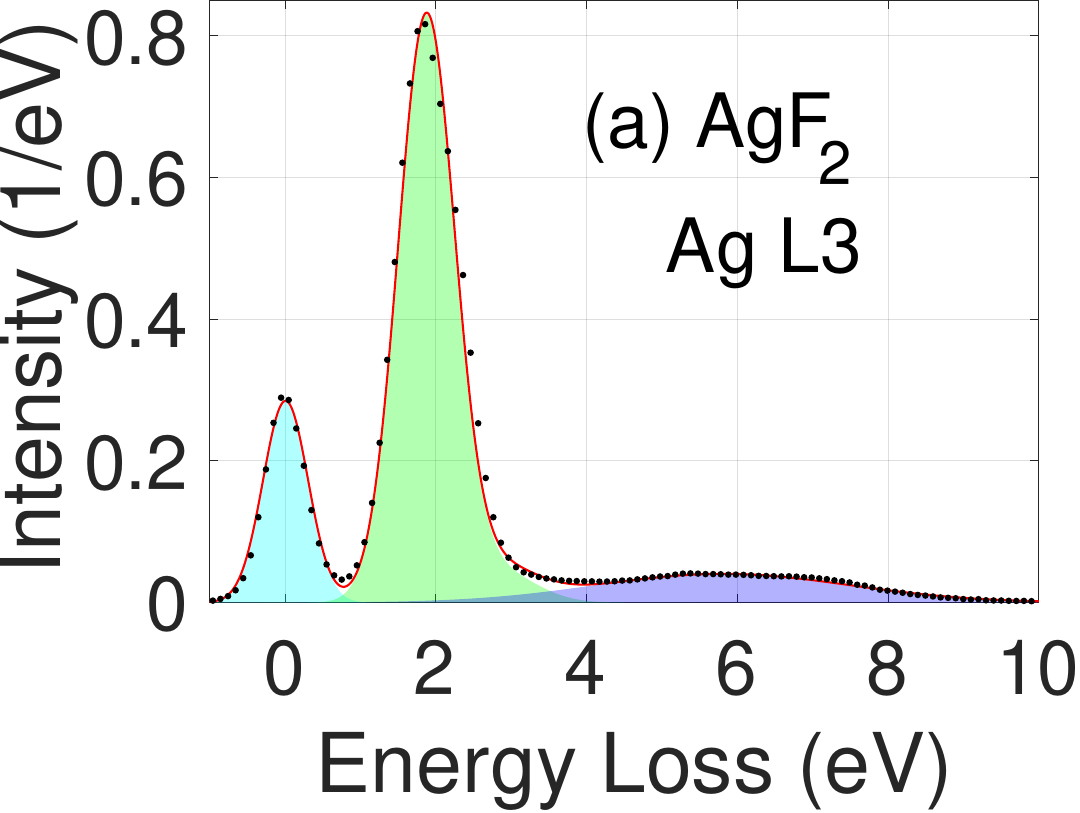}
    \end{subfigure}
    \hfill    \begin{subfigure}[t]{0.48\linewidth}
        \centering        \includegraphics[width=\linewidth]{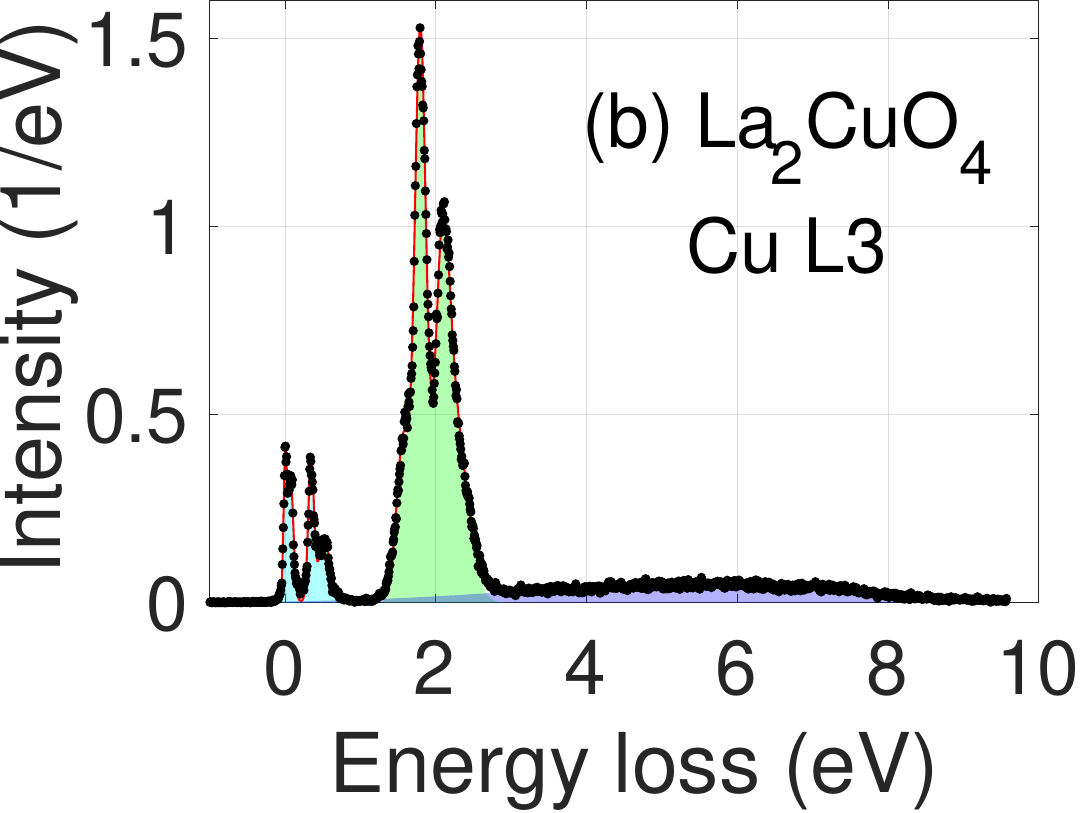}
    \end{subfigure}
    \hfill
    \begin{subfigure}[t]{0.48\linewidth}
        \centering        \includegraphics[width=\linewidth]{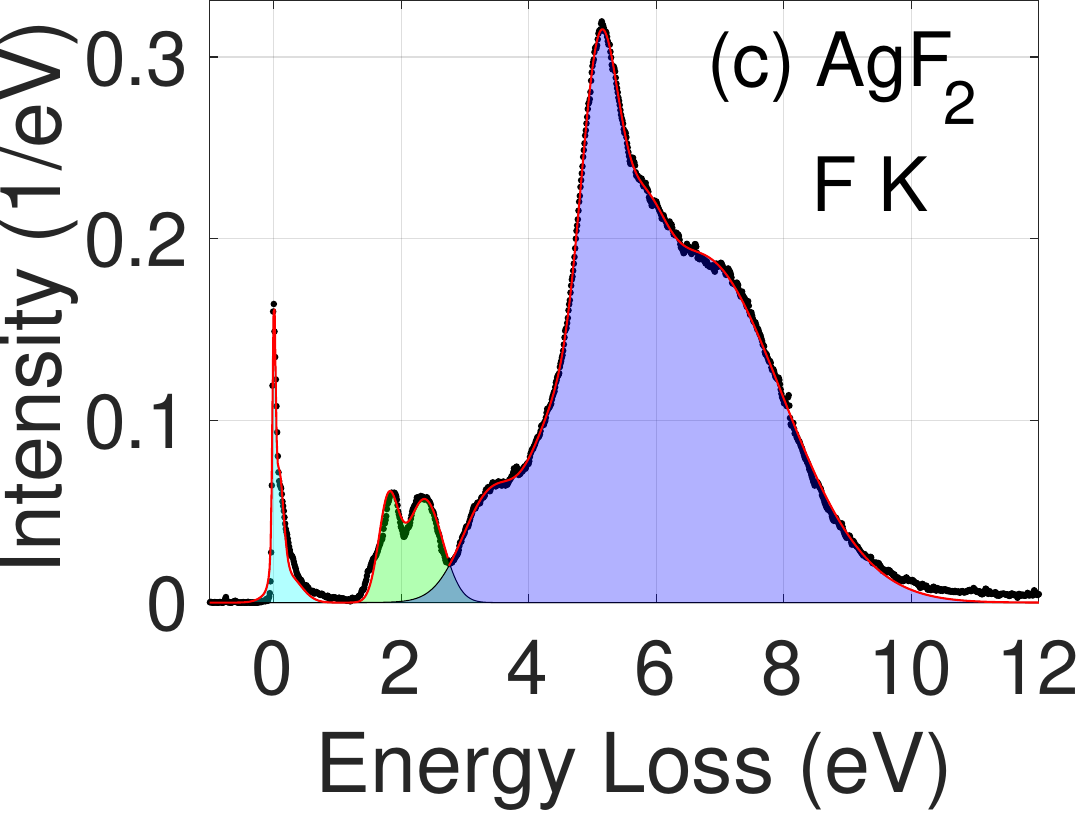}
    \end{subfigure}
    \hfill
    \begin{subfigure}[t]{0.48\linewidth}
    \centering        \includegraphics[width=\linewidth]
    {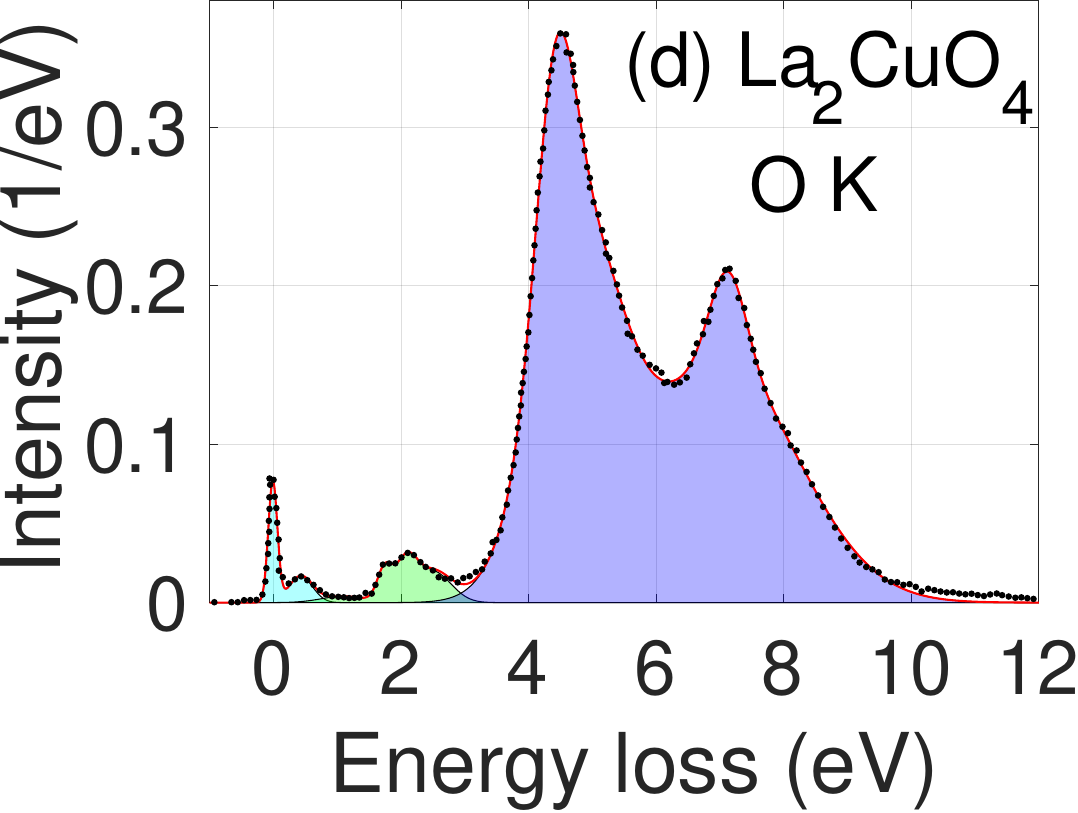}
    \end{subfigure}    
    \caption{Fits of elastic, $dd$, and CT peaks by several Gaussians for metal and ligand edges of \ch{AgF2} and \ch{La2CuO4}. Correspondent areas on the graphs are marked by various colors (light blue for quasi-elastic peak, green for $dd$ excitations, and violet for CT). Panel (a) represents the results of the current experiment, while panels (b),(c),(d) are based on the previously measured data. All the curves and fits are normalized over $dd$+CT area}
    \label{fig:fitting_L3}
\end{figure}

Fig.~\ref{fig:fitting_L3} shows the fit of the RIXS excitation spectra defining the spectral weights shown in Table~\ref{tab:dd_CT_int}.  

\begin{table*}%
\caption{\label{tab:weights} Weights of transitions from a
$\ket{d^9,x^2-y^2\downarrow}$ to possible ionic final $d$ states in terms of the incoming ($\bm\epsilon$) and outgoing  ($\bm\epsilon'$) polarization vector (first column after the line) and for $\pi$ and $\sigma$ incoming polarizations summing over outgoing polarizations. Polarization vectors are assumed to be real.  We sum over all spin directions, so the elastic and the spin-flip weights are contained in the first row. The last line is the sum over the columns. 
For simplicity, the weights are multiplied by the factors indicated in the first row. The absolute weights correspond to the result for states constructed in terms of normalized spherical harmonics and omitting radial wave functions.}
\[\arraycolsep=1.4pt\def\arraystretch{1.3}
\begin{array}{c|ccccc}
\hline\hline
\nu&\ \ \ \ \ \ \ \ 225 W_\nu&225 W_\nu(\pi)/\cos^2{\theta}&225 W_\nu(\sigma)&375A_\nu&375B_\nu\\
\hline
x^2-y^2& ({\hat \epsilon_x} {\hat \epsilon_y'}-{\hat \epsilon_x'} {\hat \epsilon_y})^2+4 ({\hat \epsilon_x'} {\hat \epsilon_x}+{\hat \epsilon_y'} {\hat \epsilon_y})^2& 3+2 \cos(2\theta')&\frac12[9+ \cos(2\theta')]&{59}/{18}&{23}/{18}\\
z^2& \frac{1}{3} 
   \left\{ (\hat \epsilon_x'{\hat \epsilon_y})^2+ (\hat \epsilon_x\hat \epsilon_y')^2-6 {\hat \epsilon_x'} {\hat \epsilon_x} {\hat \epsilon_y'} {\hat \epsilon_y}\right.& \frac{5}{3} &\frac16[13-3 \cos(2\theta')]&{121}/{54}&{-1}/{18} \\
   &\left.+   4 {\hat \epsilon_x}^2 \left[(\hat \epsilon_x')^2+(\hat \epsilon_z')^2\right]
         +4 {\hat \epsilon_y}^2 \left[(\hat \epsilon_y')^2+(\hat \epsilon_z')^2\right]\right\}&\\
xy& 4( {\hat \epsilon_x} {\hat \epsilon_y'}- {\hat \epsilon_x'} {\hat \epsilon_y})^2+({\hat \epsilon_x'} {\hat \epsilon_x}+{\hat \epsilon_y'} {\hat \epsilon_y})^2 &\frac{1}{2} [9+\cos (2\theta')]& 3+2\cos (2\theta')&{71}/{18}&{-13}/{18}\\
yz & \left({\hat \epsilon_x}^2+{\hat \epsilon_y}^2\right)\left[(\hat \epsilon_y')^2+(\hat \epsilon_z')^2\right]+3 (\hat \epsilon_y   \epsilon_z')^2
& \frac{1}{2} [3-\cos (2 \theta')]& 3-2\cos (2\theta')&{8}/{3}&-2/9\\
xz& \left({\hat \epsilon_x}^2+{\hat \epsilon_y}^2\right)[(\hat \epsilon_x')^2 +(\hat \epsilon_z')^2]+3 (\hat \epsilon_x \hat \epsilon_z')^2  &\frac{1}{2} [5-3 \cos (2\theta')]&1&{8}/{3}&-2/9\\[5pt]
\text{Total}&\frac{19}{3} \left\{\hat\epsilon_y^2 \left[(\hat\epsilon_x')^2+(\hat\epsilon_z')^2 \right]+{\hat\epsilon_x}^2 \left[(\hat\epsilon_y')^2+
(\hat\epsilon_z')^2\right]\right\}&\frac{79}{6}+\frac12\cos (2\theta')&{41}/{3}&799/54&1/18\\
&+\frac{22}{3} \left[(\hat\epsilon_x')^2 {\hat\epsilon_x}^2+(\hat\epsilon_y')^2 {\hat\epsilon_y}^2\right]-2 \hat\epsilon_x'
   {\hat\epsilon_x} {\hat\epsilon_y'} {\hat\epsilon_y}&\\
\hline\hline
\end{array}
\]
\end{table*}

\section{Weights of transitions}\label{app:weights}
\subsection{Oriented samples}
If the outgoing polarization is determined, the sum over polarization should not be done, and the weights read, 
\begin{equation}
W_\nu=\sum_{\sigma}
|\bra{d^9,x^2-y^2\downarrow}(\hat{\bm\epsilon}.\hat{\bm r}) P_{3/2}(\hat{\bm\epsilon}'.\hat{\bm r}) \ket{d^9,\nu\sigma}|^2
\end{equation}
where we assumed real polarization vectors. 
A more general treatment allowing for circularly polarized radiation has been recently presented in Ref.~\cite{Tagliavini2025}.

Table \ref{tab:weights} shows the weights for the transitions from the $x^2-y^2 \downarrow$ state to the possible $d$ states for fixed linear polarizations.

It is convenient to define a complex scattering tensor,
\begin{equation}\label{eq:T}
T_{ij}^{\nu\sigma}=\bra{d^9,x^2-y^2\downarrow}\hat{ r}_i P_{3/2}\hat{r}_j \ket{d^9,\nu\sigma}
\end{equation}

The weight can be written as,
\begin{equation}\label{eq:iddt}
W_\nu=\sum_{\sigma}
|\hat{\epsilon}_i T_{ij}^{\nu\sigma}\hat{\epsilon}'_{ j}|^2
\end{equation}
where the sum over repeated indices is understood.

If the outgoing polarization is not measured, one needs to sum over transverse polarizations $\alpha=\pi,\sigma$, which can be taken care of by introducing the projector into the transverse outgoing polarizations $P_{ij}'$:
\begin{equation}\label{eq:idd2sum}
W_\nu=\sum_{\sigma\alpha}
|\hat{\epsilon}_i T_{ij}^{\nu\sigma}\hat{\epsilon}'_{ \alpha j}|^2=\sum_{\sigma}
\hat{\epsilon}_i T_{ik}^{\nu\sigma}P_{kl}'
(T_{jl}^{\nu\sigma})^* \hat{\epsilon}_j 
\end{equation}
with   $P_{ij}'\equiv\sum_{\alpha}\hat{\epsilon}'_{ \alpha i}\hat{\epsilon}'_{ \alpha j}=\delta_{ij}-\hat k_i'\hat k_j'$ and $\hat {\bm k}_i'$ is the outgoing wavevector versor.  Table~\ref{tab:weights} and Fig.~\ref{fig:wdtsigpi} shows the weights for general polarizations and 
in the case of incoming $\pi$, $\sigma$ polarization [$W_\nu(\pi/\sigma)$] and for $\hat{\bm k}$ ($\hat{\bm k'}$)  in the $x,z$ plane and forming an angle $\theta$, ($\theta'$) with the $z$ axis. i.e. 
$\hat{\bm k}=(\sin\theta,0,\cos\theta)$,
$\hat{\bm k}'=(\sin\theta',0,\cos\theta')$. Interestingly, the sum of all weights is independent of the angles in the case of initial $\sigma$ polarization and has a very weak dependence on the outgoing angle in the case of $\pi$ incident polarization (keeping a $\cos^2\theta$ dependence). 

\subsection{Symmetry constraints}
\begin{figure}[htb]
    \centering
    \includegraphics[width=\linewidth]{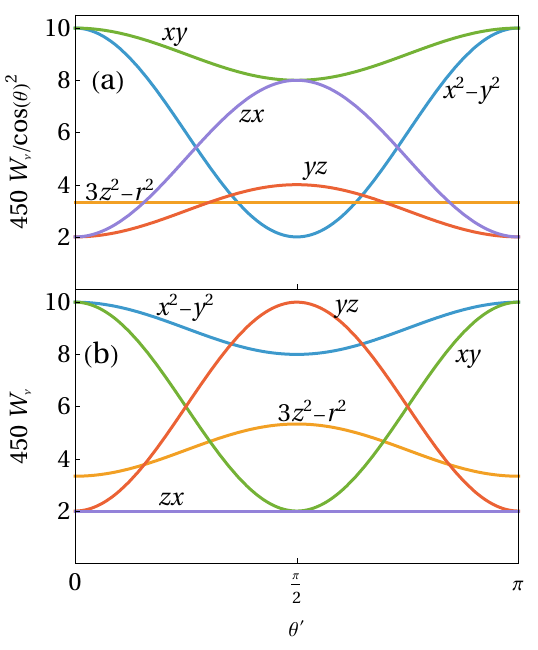}
    \caption{Weight of the transitions for different final states as a function of the outgoing scattering angle $\theta'$ for $\pi$ (a) and $\sigma$ (b) incoming polarization.}
    \label{fig:wdtsigpi}
\end{figure}

For $\theta=\theta'=0$, any plane perpendicular to $\hat{\bm k}$ can be considered as the scattering plane and therefore $\sigma$ and $\pi$ polarization become equivalent. Indeed, one can verify that in this case, the two corresponding columns in Table~\ref{tab:weights} coincide, as also shown in Fig.~\ref{fig:wdtsigpi} at the origin. Irrespectively of the polarization, also the $xy$ and the $x^2-y^2$ polarization should be equal for $\theta'=0$ as the orbitals are related by a $\pi/4$ rotation around the $z$ axis. 

\begin{table}[htb]
    \caption{\label{tab:param} Crystal fields and hybridizations used for \ch{AgF2} and LCO. All values are in eV. The diagonal energies of $P$-orbitals in Eq.~\ref{diagonal} are determined  by $\varepsilon_P^\nu=\Delta+e_P^\nu$ with the reference value $\Delta_{\ch{AgF2}}=1.29$ eV corresponding to the ``Local" parameter set of Ref.~\cite{Bachar2022}. The last column shows the expressions for the hybridizations in terms of the $x^2-y^2$ matrix element and for a planar $D_{4h}$ cluster used by Eskes {\it et al.}\cite{Eskes1990}. Columns with an asterisk are the parameters that were rescaled in the same proportion to vary the degree of covalency in Fig.~\ref{fig:wdtsigpi}.}
    \begin{ruledtabular}
    \begin{tabular}{c|ccc|ccc}
     &   & AgF$_2$    &  &    & La$_2$CuO$_4$  & \\  
        $\nu$       & $\varepsilon_d^\nu$ & $e_P^{\nu*}$        & {$T_{pd}^{\nu*}$} & $\varepsilon_d^\nu$ & $e_P^{\nu*}$        & $T_{pd}^{\nu*}$                           \\   \hline       
               $x^2-y^2$   & -0.28             & -0.16             & 2.76          & 0                 & $-\frac65 T_{pp}$ & ${T_{pd}}$                        \\
        $3z^2-r^3$       & -0.25             & 0.32              & 1.51          & 0                 & $\frac45 T_{pp}$  & $\frac{1}{\sqrt{3}}T_{pd}$        \\
        $x y$       & 0.34              & -0.05             & 1.36          & 0                 & $\frac45 T_{pp}$  & $\frac{1}{2}{T_{pd}}$             \\
        $x z$       & 0.09              & -0.14             & 1.05          & 0                 & $-\frac15 T_{pp}$ & $\frac{1}{2 \sqrt{2}}{T_{pd}}$    \\
        $y z$       & 0.10              & 0.04              & 1.02          & 0                 & $-\frac15 T_{pp}$ & $\frac{1}{2 \sqrt{2}}{T_{pd}}$    \\
    \end{tabular}
    \end{ruledtabular}
    \vspace{1ex}
\end{table}

\subsection{Arbitrarily oriented samples and  powder average}
The tensors $T_{ij}^{\nu\sigma}$ are given in the crystal reference frame. For an arbitrary orientation of the crystal, we can define a rotated tensor, 
$$
T_{ab}^{\nu\sigma}(R)=R_{ai}T_{ij}R_{bj}
$$
where the rotation matrix $R$ depends on the Euler angles $\alpha,\beta,\gamma$. Replacing in Eq.~\eqref{eq:idd2sum} yields the result for a crystal with an arbitrary orientation. For a powder, we need to integrate over all possible orientations, 
$$
\expval{W_\nu}= \sum_{\sigma}\int dR
\hat{\epsilon}_a R_{ai}  T_{ik}^{\nu\sigma}R_{bk} P_{bc}'R_{cl}
(T_{jl}^{\nu\sigma})^* R_{dj} \hat{\epsilon}_d 
$$
with the Haar measure defined as, 
$$\int dR\equiv \frac1{8\pi^2}\int_0^{2\pi} d\alpha \int_0^{\pi}d\beta\int_0^{2\pi}d\gamma \sin\beta .$$
The integrals can be done for a generic scattering matrix $T^{\nu\sigma}$,
\begin{equation}\label{eq:wpowder}
    \begin{split}
    \expval{W_\nu}
    &=\\
       \frac1{60}&\sum_\sigma\{14 \Tr(T^{\nu\sigma}T^{\nu\sigma\dagger})
-       \Tr(T^{\nu\sigma}T^{\nu\sigma*})
-       \abs{\Tr T^{\nu\sigma}}^2      \\        
     &+[ 3\abs{\Tr T^{\nu\sigma}}^2 +3  \Tr(T^{\nu\sigma}T^{\nu\sigma*})-2 \Tr(T^{\nu\sigma}T^{\nu\sigma\dagger})]S \} \\&
     =A_{ \nu }  + B_{\nu} S
    \end{split}
\end{equation}
with $S=\cos[2(\theta-\theta')]$ for incoming $\pi$ polarization and $S=1$ for  incoming $\sigma$ polarization. %
Table~\ref{tab:weights} shows the constants $A_{ \nu }$ $B_{\nu}$ in the present case.

Numerical weights for notable polarizations applicable for oriented crystals and powders are given in Table~\ref{tab:rweights} of the main text.

\section{Model and Parameters}\label{app:param}
For the calculations,  we consider a cluster with a central metal atom and four neighbouring ligands in the case of LCO and six ligands in the case of \ch{AgF2}. In the case of LCO, we neglect the apical oxygen atoms, which, as shown in Ref.~\cite{Eskes1990}, is a good first approximation. This provides a reference model with a minimal parameter set. 
Considering the Hamiltonian in terms of holes and in the case of one hole, interactions are irrelevant and the Hamiltonian separates into five $2\times2$ problems, one for each sector,
\begin{eqnarray}
    \label{diagonal}
    H &=& \sum_{\nu\sigma} \varepsilon_{d}^\nu d^\dagger_{\nu\sigma}  d_{\nu\sigma} +\sum_\nu \varepsilon_P^\nu P^\dagger_{\nu\sigma}  P_{\nu\sigma} \nonumber\\
    &+& \sum_{\nu\sigma} T_{pd}^\nu\left( {d}^\dagger_{\nu\sigma}  P_{\nu\sigma} +  P^\dagger_{\nu\sigma}  d_{\nu\sigma} \right). 
\end{eqnarray}
Here $ d^\dagger_{\nu\sigma}$ creates a hole in the $d$-orbitals ($\nu={x^2-y^2},{3z^2-r^2}, {xy},  {xz}, {yz}$) with spin $\sigma$ while $P^\dagger_{\nu\sigma}$ creates a hole in a linear combination of the $p$ orbitals of the ligand which transform like one of the $d$ orbitals. For details see Refs.~\cite{Eskes1990,Piombo2022,Bachar2022}. Diagonal energies and hybridizations in the Hamiltonian are specified in Table~\ref{tab:param}.
We define $T_{pd}\equiv T_{pd}^{x^2-y^2}$.

Cuprates parameters are from Ref.~\cite{Eskes1990} as explained in the main text. These authors neglect the crystal field splitting of the $d^9$ configuration. The splitting of the $d^{10}\underline{L}$ is entirely due to $T_{pp}$. Table~\ref{tab:param} shows the parameters used in the computations. Columns with an asterisk indicate parameters rescaled in the same proportion as $T_{pd}/\Delta$ to draw Fig.~\ref{fig:ratiolasco}.

%

\end{document}